\documentclass[usenatbib]{mnras}

\usepackage{newtxtext,newtxmath}

\usepackage[T1]{fontenc}
\usepackage{ae,aecompl}

\usepackage{graphicx}	
\usepackage{amsmath}	
\usepackage{amstext}
\usepackage[usenames]{xcolor}
\usepackage{booktabs,gensymb}
\usepackage{comment}

\newcommand{\revis}[1]{#1}
\newcommand{\revi}[1]{#1}
\newcommand{\rev}[1]{#1}

\newcommand{\Msun}{\,M_{\odot}}
\newcommand{\Mh}{M_{\mathrm{h}}}
\newcommand{\Ms}{M_{*}}
\newcommand{\Rvir}{R_{\mathrm{vir}}}
\newcommand{\epsff}{\epsilon_{\mathrm{ff}}}
\newcommand{\lcrit}{\lambda_{\mathrm{crit}}} 
\newcommand{\sigmacell}{\sigma_{\mathrm{cell}}}
\newcommand{\tdep}{t_{\mathrm{dep}}}
\newcommand{\sfr}{\mathrm{SFR}}
\newcommand{\Hm}{{\mathrm{H}_2}}
\newcommand{\HI}{{\mathrm{HI}}}
\newcommand{\Hn}{{\mathrm{HI+H}_2}}
\newcommand{\Mpcs}{\Msun\,\mathrm{pc}^{-2}}

\title[Structure of high-redshift galaxies]{Structure and stability of high-redshift galaxies in cosmological simulations}

\author[Meng et al.]{
Xi Meng$^{1}$\thanks{E-mail: xim@umich.edu}\href{https://orcid.org/0000-0002-8276-4164}{\includegraphics[scale=0.6]{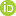}},
Oleg Y. Gnedin$^{1}$\href{https://orcid.org/0000-0001-9852-9954}{\includegraphics[scale=0.6]{orcid.png}}
and Hui Li$^{2}$\href{https://orcid.org/0000-0002-1253-2763}{\includegraphics[scale=0.6]{orcid.png}}
\\
$^{1}$Department of Astronomy, University of Michigan, Ann Arbor, MI 48109, USA\\
$^{2}$Department of Physics, Kavli Institute for Astrophysics and Space Research, MIT, Cambridge, MA 02139, USA
}

\date{Accepted XXX. Received YYY; in original form ZZZ}

\pubyear{2018}

\begin{document}
\label{firstpage}
\pagerange{\pageref{firstpage}--\pageref{lastpage}}
\maketitle

\begin{abstract}
We investigate the structure of galaxies formed in a suite of high-resolution cosmological simulations. Consistent with observations of high-redshift galaxies, \revi{our simulated galaxies show irregular, prolate shapes}, which are dominated by turbulent motions instead of rotation. Yet molecular gas and young stars are restricted to \revis{a relatively thin plane}. We examine the accuracy of applying the Toomre linear stability analysis to predict the location and amount of gas available for star formation. We find that the Toomre criterion still works for these irregular galaxies, after correcting for multiple gas and stellar components: the $Q$ parameter in $\Hm$ rich regions is in the range $0.5-1$, remarkably close to unity. Due to the violent stellar feedback from supernovae and strong turbulent motions, young stars and molecular gas are not always spatially associated. Neither the $Q$ map nor the $\Hm$ surface density map coincide with recent star formation exactly. We argue that the Toomre criterion is a better indicator of future star formation than a single $\Hm$ surface density threshold because of the smaller dynamic range of $Q$. The depletion time of molecular gas is below 1~Gyr on kpc scale, but with large scatter. Centering the aperture on density peaks of gas/young stars systematically biases the depletion time to larger/smaller values and increases the scatter. 
\end{abstract}




\section{Introduction}

Recent analytical models of star formation (SF) in galaxies \citep[e.g.,][]{Dekel:2009aa,Vollmer:2011aa,Kruijssen:2014aa,Krumholz:2018aa} have used the Toomre criterion $Q\lesssim 1$ to identify regions of dense gas in \rev{disc} galaxies that should collapse and form stars. The original Toomre analysis \citep{toomre64} considers stability of an isothermal gas layer to linear axisymmetric perturbations in regular, thin, axisymmetric \rev{discs}. These conditions are not strictly satisfied even in many galaxies at low redshift, and are certainly violated in clumpy, turbulent \rev{disc} galaxies at high redshift. Without invoking the Toomre criterion, however, it is difficult to construct predictive models that could help interpret observations of galaxies at $z\gtrsim2$ that are expected from powerful oncoming and future facilities. In this paper we test the validity of the Toomre analysis for turbulent irregular galaxies using new state-of-the-art simulations of galaxy formation.

Deep HST-UDF observations indicate that normal spiral galaxies appear only at redshifts $z<1.5$ \citep{Elmegreen:2014aa}. At higher redshifts galaxies tend to have thick stellar \rev{discs} fragmented into dense stellar clumps \citep[e.g.,][]{Genzel:2010aa, Elmegreen:2017aa}. These galaxies have kinematics dominated by turbulent motions and show comparable amounts of molecular gas and stars \citep{Tacconi:2013aa}. In some cases enhanced spatial resolution afforded by gravitational lensing reveals that the locations molecular gas and UV emission from young stars are spatially decoupled on sub-kpc scales \citep{Dessauges-Zavadsky:2017aa}. Measurements of the Toomre $Q$ parameter find values exceeding unity for ionized H$\alpha$ gas but giant clumps appear to be unstable with $Q<1$ \citep{Genzel:2014aa}. Similarly, in a study of low-redshift analogues of turbulent disk galaxies \citet{Fisher:2017aa} find that large H$\alpha$ clumps exist only in regions with $Q<1$.

Numerical simulations have also studied stability of galactic \rev{discs}. \citet{Li:2005aa, Li:2006aa} calculated the Toomre $Q$ for a combination of collisional gas and collisionless stars in isolated \rev{disc} simulations. \citet{Ceverino:2010aa} and \citet{Inoue:2016aa} used cosmological simulations of clumpy high-redshift galaxies and found values of $Q>1$ in large parts of the \rev{discs}. More recent isolated galaxy simulations show that stellar clumps fragment further on smaller sub-kpc scales, below the characteristic Toomre mass \citep{Tamburello:2015aa}. These simulations of high-redshift galaxies show significantly irregular structure of the \rev{discs} due to fast gas accretion and stellar feedback. The kinematics of the interstellar medium includes strong turbulent motions on all scales.

Given this complexity, is it correct to apply the Toomre analysis to select regions undergoing gravitational collapse and star formation? How well can we predict the amount of star-forming gas and star formation rate (SFR) in high-redshift galaxies? 

Our goal is to estimate the accuracy of applying the Toomre criterion and Toomre fragmentation mass. Using a recent suite of cosmological simulations of Milky Way-type galaxies, we verify the validity of the Toomre analysis. \rev{Unlike regular disc galaxies observed at low redshift, our simulated high redshift galaxies are turbulent and show irregular shapes. To explore the structure of our simulated galaxies, we study the radial profiles of these galaxies and use moments of inertia to obtain their axis ratios in order to check whether they can be considered discs.} We argue that the $Q$ parameter is a better predictor of the star formation sites than simply the molecular gas density, because it covers a smaller dynamic range. We also study the depletion time of cold gas as a function of spatial scale and, in agreement with previous non-cosmological studies, show that the two commonly used approaches (gas-centred and star-centred) give very different estimates on $\sim 100$~pc scales but converge on kpc scale.

We describe the simulations used for this analysis and the spatial and kinematic structure of simulated galaxies in Section~\ref{sec:sim}. We apply the Toomre analysis and calculate maps of the $Q$ parameter in Section~\ref{sec:toomre}. In Section~\ref{sec:depletion} we study the depletion time of cold molecular gas on different spatial scales. We discuss the implications of our results for modeling galactic star formation in Section~\ref{sec:discussion} and present our conclusions in Section~\ref{sec:summary}.

\begin{table*}
\centering
\caption{Global properties of the main galaxy} \label{tab:basic}
\begin{tabular}{lccccccccc}
\toprule
Run & $z$ & $\Mh\,(\Msun)$ & $\Ms\,(\Msun)$ & $M_{\Hn}\,(\Msun)$ & $M_{\Hm}\,(\Msun)$ & $\Rvir\,$(kpc) & $R_{\rm h,*}\,$(kpc) & $R_{\rm h,\Hn}\,$(kpc) & $R_{\rm h,\Hm}\,$(kpc)\\
     \midrule
SFE200     & 1.78 & $2.37 \times 10^{11}$ & $4.06 \times 10^9$ & $2.65\times 10^9$ & $1.77\times 10^8$ & 67.9 & 2.6 & 3.2 & 2.3\\     
SFE100     & 1.50 & $4.36 \times 10^{11}$ & $8.19 \times 10^9$ & $7.23\times 10^9$ & $8.54\times 10^8$ & 91.5 & 2.7 & 4.7 & 2.2\\
SFE50      & 1.50 & $4.35 \times 10^{11}$ & $7.13 \times 10^9$ & $8.35\times 10^9$ & $7.47\times 10^8$ & 91.5 & 4.0 & 8.6 & 2.8\\
SFE10      & 1.78 & $2.50 \times 10^{11}$ & $6.46 \times 10^9$ & $4.98\times 10^9$ & $1.15\times 10^9$ & 69.0 & 2.4 & 3.6 & 2.3\\
SFEturb    & 2.85 & $1.21 \times 10^{11}$ & $2.44 \times 10^9$ & $2.78\times 10^9$ & $2.59\times 10^8$ & 39.2 & 2.2 & 4.0 & 2.1\\
SFE50-SNR3 & 1.98 & $2.40 \times 10^{11}$ & $1.37 \times 10^{10}$ & $3.29\times 10^9$ & $1.88\times 10^9$ & 63.5 & 0.7 & 1.3 & 0.9\\
     \bottomrule
\end{tabular}
\end{table*}

\section{Structure of high redshift galaxies} \label{sec:sim}

\subsection{Simulation Suite}

We use a suite of cosmological simulations described in \citet{li_etal17,li_etal18}. These simulations were run with the Adaptive Refinement Tree (ART) code \citep{Kravtsov:1997aa,Kravtsov:1999aa,Kravtsov:2003aa,Rudd:2008aa} in a periodic box of 4 comoving Mpc. All runs start with the same initial conditions but use different sub-grid model parameters of star formation and stellar feedback. The initial conditions were selected to produce a main halo with total mass $M_{200}\approx 10^{12}\Msun$ at $z=0$, similar to that of the Milky Way. The ART code uses adaptive mesh refinement to increase spatial resolution in the dense galactic regions. The lowest resolution level is set by the root grid, which in these runs was $128^3$ cells. This sets the dark matter particle mass $m_{\rm DM}=1.05\times 10^6\Msun$. The refinement strategy is quasi-Lagrangian, which keeps cell mass within a narrow range. The finest refinement level is chosen so that the physical size of gas cells at that level is between 3 and 6~pc. This required increasing the number of additional refinement levels gradually with time, from 9 levels at $z>9$ to 10, 11, and 12 refinement levels at $z\approx 9,\,4,\,1.5$, respectively. 
 
The simulations include three-dimensional radiative transfer using the Optically Thin Variable Eddington Tensor approximation \citep{Gnedin:2001aa} of ionizing and ultraviolet radiation from stars \citep{Gnedin:2014aa} and the extragalactic UV background \citep{Haardt:2001aa}, non-equilibrium chemical network that deals with ionization states of hydrogen and helium, and phenomenological molecular hydrogen formation and destruction \citep{Gnedin:2011aa}. The simulations also incorporate a subgrid-scale (SGS) model for unresolved gas turbulence \citep{Schmidt:2014aa,semenov_etal16}. The star formation is implemented with a new method that follows the formation of individual star clusters. In this continuous cluster formation (CCF) algorithm \citep{li_etal17,li_etal18} each star particle represents a star cluster that forms at a local density peak and grows mass via accretion within a spherical region of fixed physical size, until the feedback of young stars terminates the growth of the star cluster. 

In addition to the early radiative and stellar wind feedback \rev{(similar to that described in \citealt{agertz_etal13})}, the simulations include a supernova (SN) remnant feedback model \citep{Martizzi:2015aa,semenov_etal16}. As the SN remnant model was calibrated by simulations of isolated SN explosion, rather than multiple SNe that appear in star clusters, its momentum feedback is underestimated. In addition, some momentum is lost due to advection errors as the SN shell moves across the simulation grid. Cosmic rays accelerated by the SN remnant could also boost the momentum deposition \citep{Diesing:2018aa}. To compensate for these effects, the momentum feedback of the SN remnant model is boosted by a factor $f_{\rm boost}$. The default value $f_{\rm boost}=5$ is chosen to match the star formation history expected from the abundance matching method, but the simulation suite contains also runs with different $f_{\rm boost}$ \rev{(see Section 2.2.8 and Figure 5 of \citealt{li_etal18})}.

\rev{We adopt a $\rm{\Lambda CDM}$ cosmology with $\Omega_m=0.304, \Omega_b=0.048, h=0.681, \sigma_8=0.829$ \citep{Planck-Collaboration:2016aa}.}

In this paper we focus on several runs with different local star formation efficiency $\epsff$ and SN momentum boost factor $f_{\rm boost}$. Table~\ref{tab:basic} contains basic information of these simulations. \rev{The last three columns in Table~\ref{tab:basic} are the radii containing half the mass of stars, neutral gas and molecular gas, respectively.} The number after "SFE" in the names corresponds to the percentage of local $\epsff$. In SFEturb run $\epsff$ is variable and turbulence-dependent \citep[as implemented by][]{semenov_etal16}. SFE50-SNR3 run is a weaker feedback run, with the lower SN boost factor $f_{\rm boost}=3$. All other runs have $f_{\rm boost}=5$. We focus on the main galaxy in the last available output of each run and list their masses and sizes in Table~\ref{tab:basic}. For run SFE200 we analyze the snapshot at $z=1.78$ (same as for SFE10 run), because at the last output ($z=1.44$) the main galaxy is experiencing a major merger and its morphology is strongly perturbed.

\revi{We note that because of strong stellar feedback, at these outputs there are few cells at the highest refinement level in a 10 kpc cube centered on the main galaxy. Therefore, the spatial resolution for our study is limited to about 100 pc.}

\begin{figure*}
    \centering
    \includegraphics[width=1.0\textwidth]{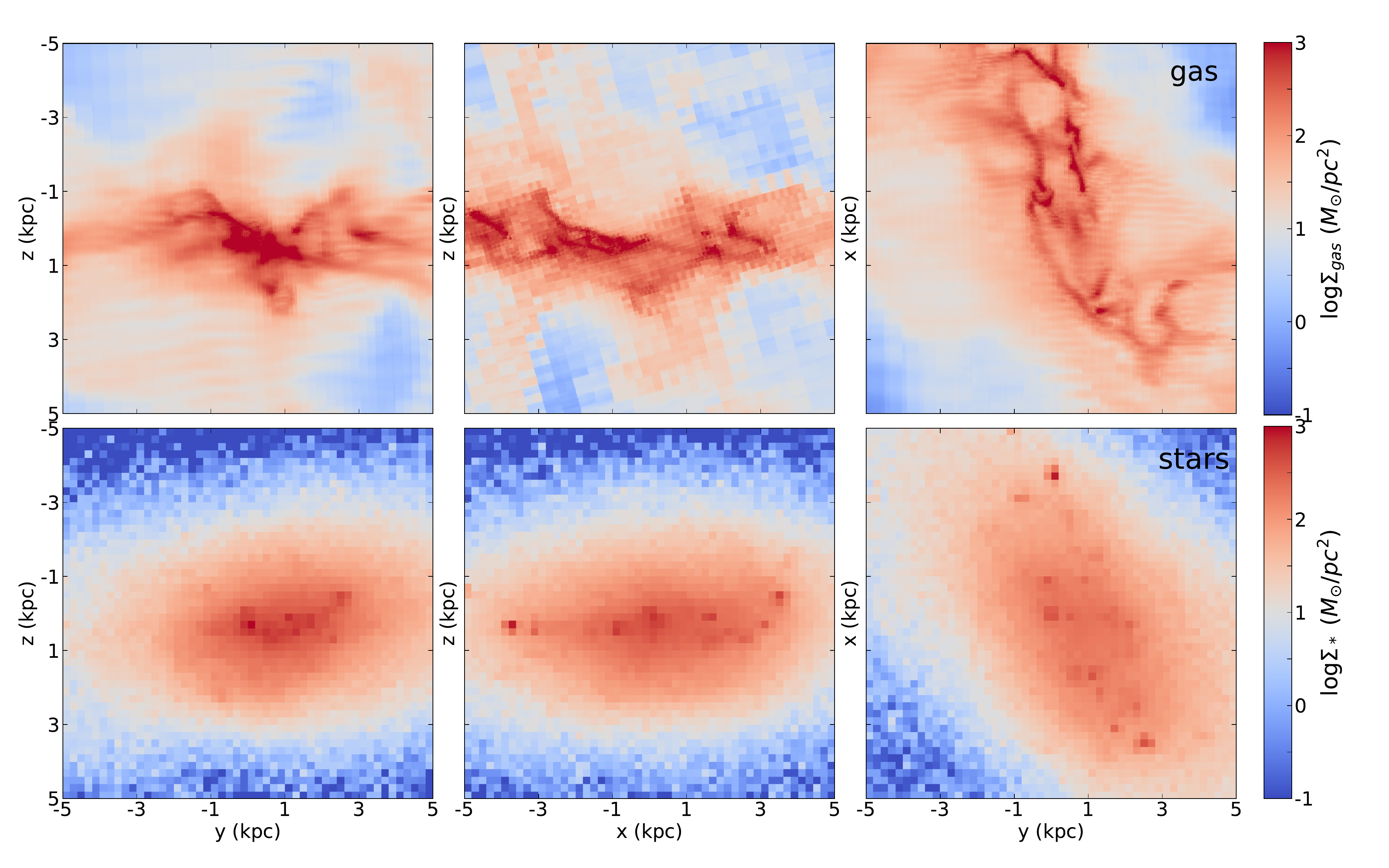}
    \vspace{-5mm}
    \caption{Density projection of all gas (upper panels) and stars (lower panels) along three principle axes given by the tensor of inertia of neutral gas, for run SFE50 at $z=1.5$. $Z$-coordinate corresponds to the minor axis. Thus the right panels are "face-on" while the left and middle panels are "edge-on" views. The projection depth is $\pm 5\,$kpc.}
     \label{fig:den_map}
\end{figure*}

\subsection{Surface density profiles}

To study the surface density profile of the simulated galaxies, we need to determine the centre of a galaxy and orientation of the projection plane. We define the galaxy centre to be at the location of maximum stellar density, which is found iteratively using smaller and smaller smoothing kernels in \citet{Brown:2018aa}. To determine the \rev{galaxy} orientation, we examined several alternative definitions based on the angular momentum of neutral gas and the principle axes of the tensor of inertia of gas and stars.
 
We use the following definition of the inertia tensor (also called "shape tensor" in \citealt{zemp_etal11})
$$ \textbf{I} \equiv \sum_{k,i,j} M_k \, r_{k,i} \, r_{k,j}\ \textbf{e}_{i} \otimes \textbf{e}_{j} $$
where $M_k$ is the mass of $k$-th stellar particle or gas cell, $r_{k,i}$ are its coordinates in the galactocentric reference frame $(i=1,2,3)$, and $\textbf{e}_i$ are the three unit vectors of the coordinate axes. The tensor can be diagonalized by a rotation matrix, to calculate the principle moments of inertia $I_1 \ge I_2 \ge I_3$. From these we calculate the axis ratios $b/a = (I_2/I_1)^{1/2}$ and $c/a = (I_3/I_1)^{1/2}$. The orientation of the \rev{galaxy plane} is given by the eigenvector corresponding to the smallest eigenvalue.

We find that the angular momentum of neutral ($\Hn$) gas enclosed within a sphere of a given radius can suddenly change direction between $0.1\,\Rvir$ and $0.3\,\Rvir$ in some runs. For example, in SFE50 run it changes by 82\degree, and in SFEturb run by 52\degree, while in the other runs is remains stable within 10\degree. At large radii the gas typically falls along cosmic filaments, whereas closer to the centre mergers of galactic clumps can significantly perturb the \rev{galaxy plane} orientation. Since most of the stars and neutral gas in the simulated galaxies are located within $0.1\,\Rvir$, the inner angular momentum is more relevant for the \rev{disc} formation. Within $0.1\,\Rvir$ the direction of the gas angular momentum is consistent to better than 20\degree\ for all runs. The direction of angular momentum of molecular $\Hm$ gas generally follows that of the neutral gas, to better than 16\degree. The scale of $0.1\,\Rvir$ corresponds to $4-9$~kpc (Table~\ref{tab:basic}); it varies from galaxy to galaxy because of the different redshift of the final available output.

Next we consider the principal axes of the shape tensor for neutral gas and stars, calculated with the $0.1\,\Rvir$ radius sphere. Their orientation can deviate from the angular momentum of neutral gas by as much as 36\degree\ in SFEturb run and 24\degree\ in SFE50 run. For the other runs they are within $\approx 10\degree$. The moments of inertia calculated separately for the gas and stars generally agree with each other.

\rev{Although we first expected that the cold gas would settle into a thin disc and its angular momentum would be the best indicator of the rotation plane, the above comparison shows that the shape tensor gives a more consistent definition of the galaxy plane. The plane given by angular momentum has some deviations from the galaxy plane identified by eye.} Therefore, we choose to use the gas shape tensor to define the \rev{galaxy} plane \rev{and use this orientation throughout the paper}. Similarly, \citet{garrison-kimmel_etal18} used the inertia tensor of stars to define the \rev{galaxy} orientation in their analysis of FIRE-2 simulations.

Figure~\ref{fig:den_map} shows examples of the density map of gas and stars projected along the three principle axes. The stellar distribution is strongly puffed up in the vertical direction and elongated in the \rev{galaxy} plane. The axis ratios from the shape tensor show that, except for SFE50-SNR3 run, our simulated galaxies are not thin, axisymmetric \rev{discs}. The axis ratios $c/a$ and $b/a$ for stars are generally around 0.4 and 0.6, respectively (for SFEturb run they are around 0.65 and 0.85), while the axis ratios for neutral gas are around 0.3 and 0.5 (for SFE200 run they are around 0.5 and 0.8). Axis ratios vary from galaxy to galaxy, but except for SFE50-SNR3 run (which has $b/a=0.9$ for stars and $b/a=0.8$ for molecular gas) they all indicate non-axisymmetric, triaxial shapes for both stars and molecular gas. This can be seen in the right panels of Figure~\ref{fig:den_map}, where the face-on view of the galaxy is not axisymmetric. 

\begin{figure}
  \centering
  \includegraphics[width=\columnwidth]{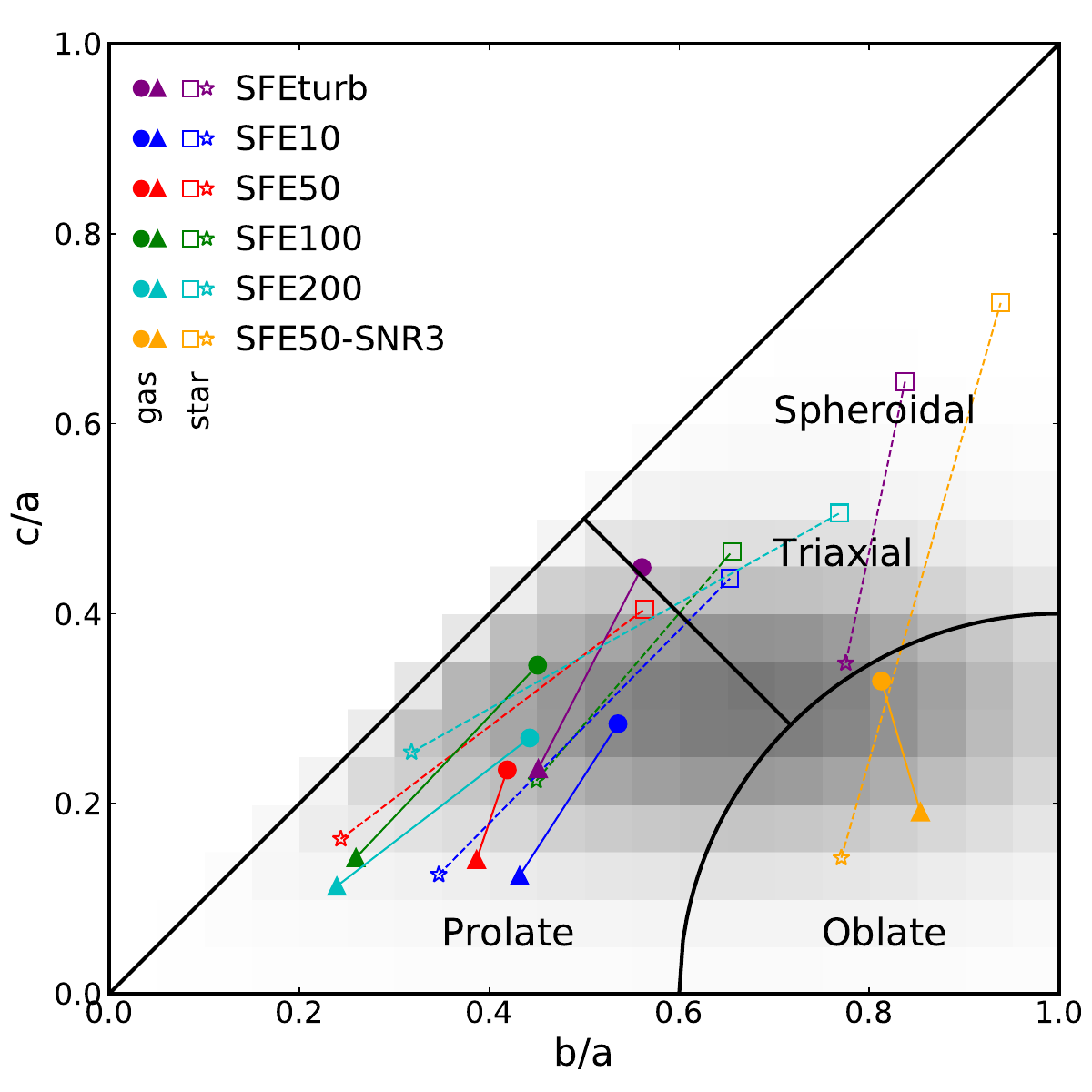}
  \vspace{-5mm}
  \caption{Axis ratios of all gas (filled circles) and molecular gas (filled triangles), and all stars (open squares) and young stars (asterisks) for different runs. The parameter space of the axis ratios $b/a$ and $c/a$ is divided into three parts to distinguish shapes of an ellipsoid, following the definition in \citet{Zhang:2019aa} and \citet{van-der-Wel:2014aa}. The gray shades show the observed distribution of axis ratios for stars in the high redshift galaxies of similar mass range to our simulated galaxies \citep{Zhang:2019aa}.}
  \label{fig:ratioshape}
\end{figure}
 
\citet{van-der-Wel:2014aa} intuitively defined galaxy shapes to be oblate, prolate, or spheroidal based on the axis ratios. Here we adopt their definition and plot the axis ratios for the gas and stars in our simulated galaxies in Figure~\ref{fig:ratioshape}. Two runs stand out from the rest. The SFE50-SNR3 run has weaker feedback, which produces a more regular \rev{disc} and shows features qualitatively different from all the other runs. The SFEturb output is at an earlier epoch, which could be the cause of its different axis ratios of stars from the other runs with the fiducial feedback strength.

As pointed out by \citet{Zhang:2019aa}, the "spheroidal" regime also contains triaxial shapes and is better named as "spheroidal or triaxial". \citet{Zhang:2019aa} assumed that the intrinsic shapes of galaxies are triaxial ellipsoids and fit ellipses with the same $b/a$ axis ratios to isophote contours (instead of using the tensor of inertia). Then they used a model to reconstruct the three-dimensional shape based on the observed two-dimensional axis ratios for galaxies in the multi-wavelength CANDELS survey. They used data in wavelengths that are as close as possible to 4600\AA\ in rest-frame, which better traces the distribution of young stars. We show in gray shades in Figure~\ref{fig:ratioshape} their modeled axis ratio distribution for the redshift bin $1.5<z<2.0$ and mass bin $9.5 < \log\Ms/\Msun < 10$, which are closest to our simulated galaxies. Compared to the observations, the distribution of stars in our galaxies is typically rounder and thicker, extending from mildly prolate to triaxial to spheroidal regimes. The shape of the gas is significantly more flattened, falling into the prolate regime. \citet{Ceverino:2015aa} and \citet{Tomassetti:2016aa} also find that the shapes of $z\approx 1-3$ galaxies in their simulations tend to be triaxial or prolate.

However, the shapes of molecular gas and young stars are much more flattened ($c/a=0.1-0.2$) and elongated ($b/a<0.5$). Since the number of young star particles is not large and their distribution is clumpy (except for SFE50-SNR3 run) it is easy for young stars and molecular gas to display an elongated, prolate configuration. The shapes of young stars and molecular gas are similar, because the stars form out of molecular gas. As the galaxy evolves and interacts with other galaxies, these newly formed stars leave their formation sites and settle into a more spheroidal configuration. 

Current observations suggest that galaxies transform from clumpy, thick \rev{discs} at $z\gtrsim 1.5$ \citep[e.g.,][]{Elmegreen:2017aa} to regular, spiral structures at low redshift as galactic \rev{discs} become less turbulent \citep{Elmegreen:2014aa}. During this process galaxy shapes transit from prolate to oblate \citep{Zhang:2019aa}, while the gas fraction decreases \citep{Dessauges-Zavadsky:2017aa}. \citet{Shibuya:2016aa} found that the fraction of clumpy galaxies increases at $z\gtrsim 2$, peaks at $z\simeq 1-2$, and decreases at $z\lesssim 1$. This also indicates that galaxies transit from irregular morphology at high redshift to more regular, disky shapes at relatively low redshift.

\begin{figure}
    \centering
\includegraphics[width=\columnwidth]{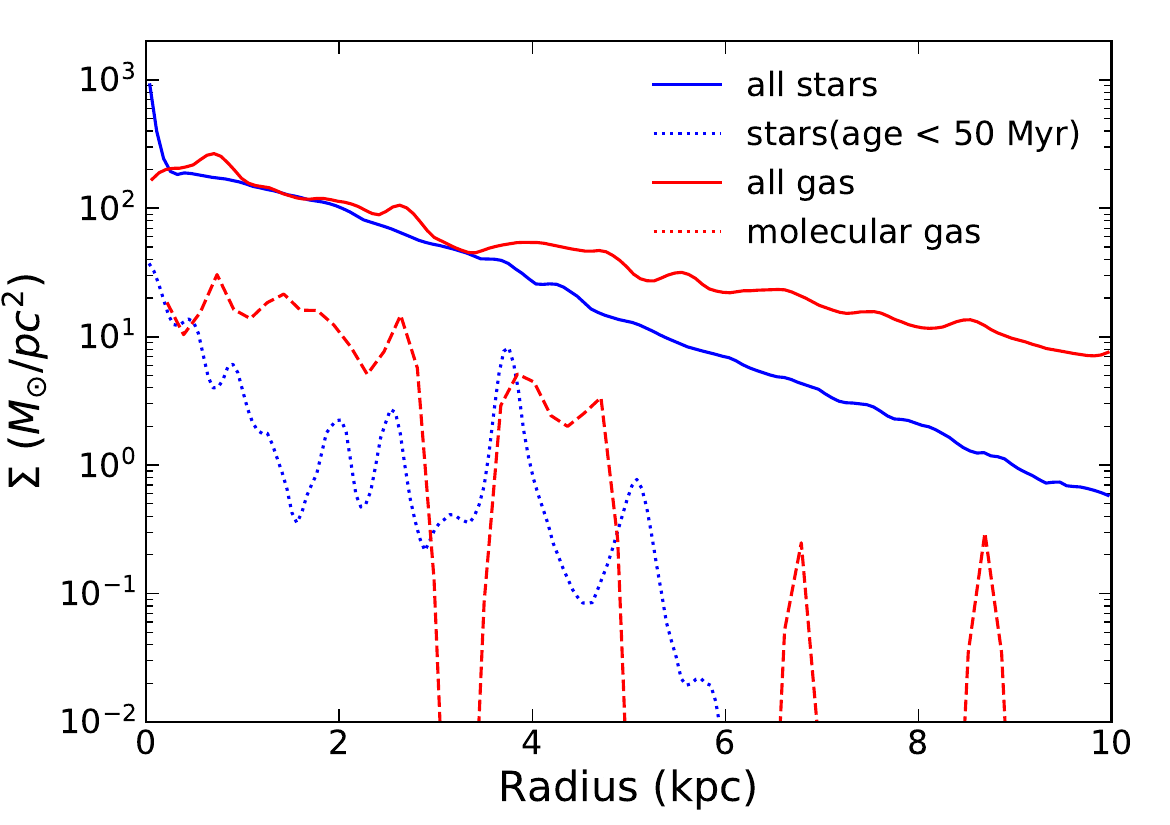}
    \vspace{-5mm}
\caption{An example of the surface density profiles of gas and stars in cylindrical shells in run SFE50. The projection depth is $\pm 5\,$kpc.}
    \label{fig:profile_radial}
\end{figure}

Figure~\ref{fig:profile_radial} shows an example of the surface density profile of stars and gas in one galaxy. \rev{The surface density profile is calculated in cylindrical shells in the galaxy plane defined previously, with projection depth being $\pm5$~kpc.} Stars dominate over gas near the centre but have similar density in the range of radii $0.4<R<3$~kpc. The gas density decreases less steeply with radius and begins to dominate over stars at larger radii. Since some of our galaxies have spheroidal shapes, we chose a large projection depth of 5~kpc to capture most of the stars and gas extending above the \rev{galaxy} plane. Because of such a deep projection the molecular gas does not dominate in any cylindrical shell, but still reaches very high column density, $10-50\Msun\,$pc$^{-2}$, in the inner few kpc. Outside $R=5$~kpc the $\Hm$ density is low. When we calculate the Toomre $Q$ parameter in later sections, we focus on the square region of $\pm5$~kpc from the centre of the galaxy \rev{in the galaxy plane with a certain thickness}, which contains most of the star-forming molecular gas. The total gas and stellar surface densities in this region are comparable. The density profile of young stars formed within the last 50~Myr is quite irregular and does not directly follow the distribution of molecular gas. We return to this point in the discussion of density maps.

\begin{figure}
    \centering
\includegraphics[width=\columnwidth]{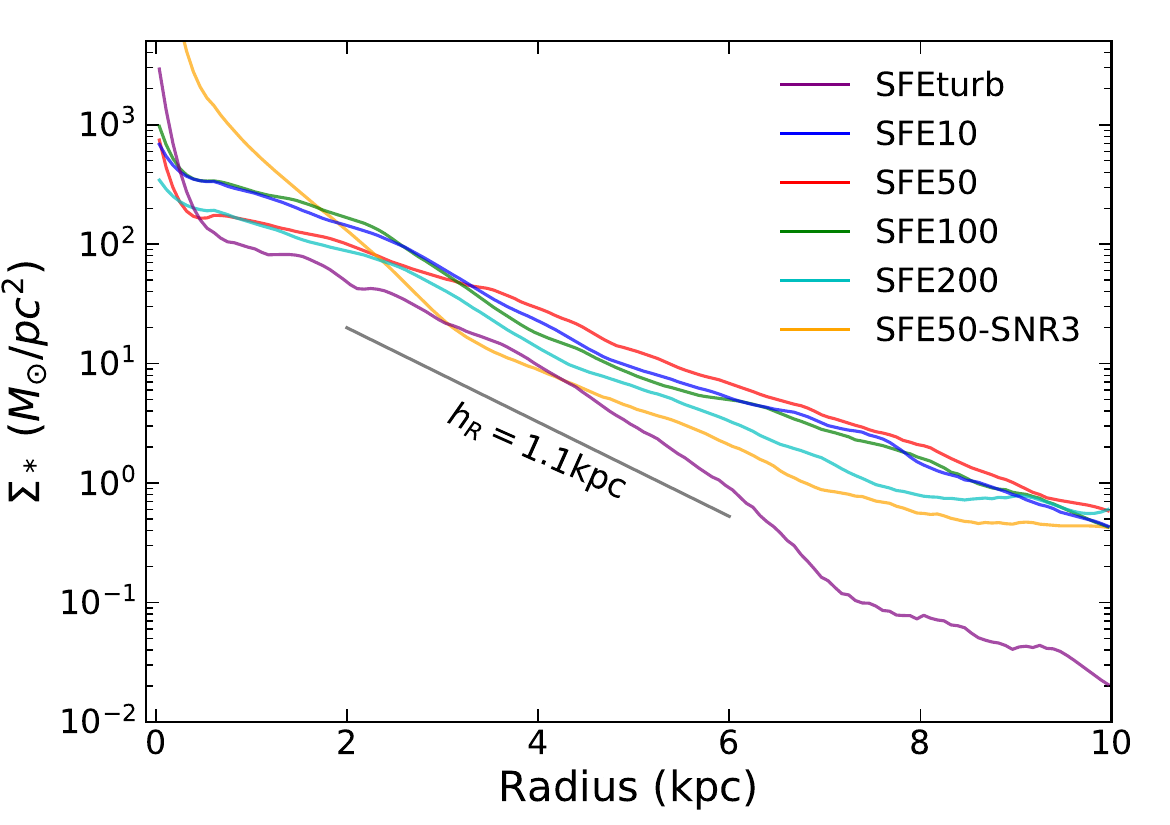}
    \vspace{-5mm}
\caption{Comparison of the stellar surface density profiles for all six runs. Gray line corresponds to an exponential \rev{disc} with the scale length of 1.1~kpc.}
	\label{fig:profile_radial_sumstar} 
\end{figure}

\begin{figure*}
    \centering
    \includegraphics[width=0.66\textwidth]{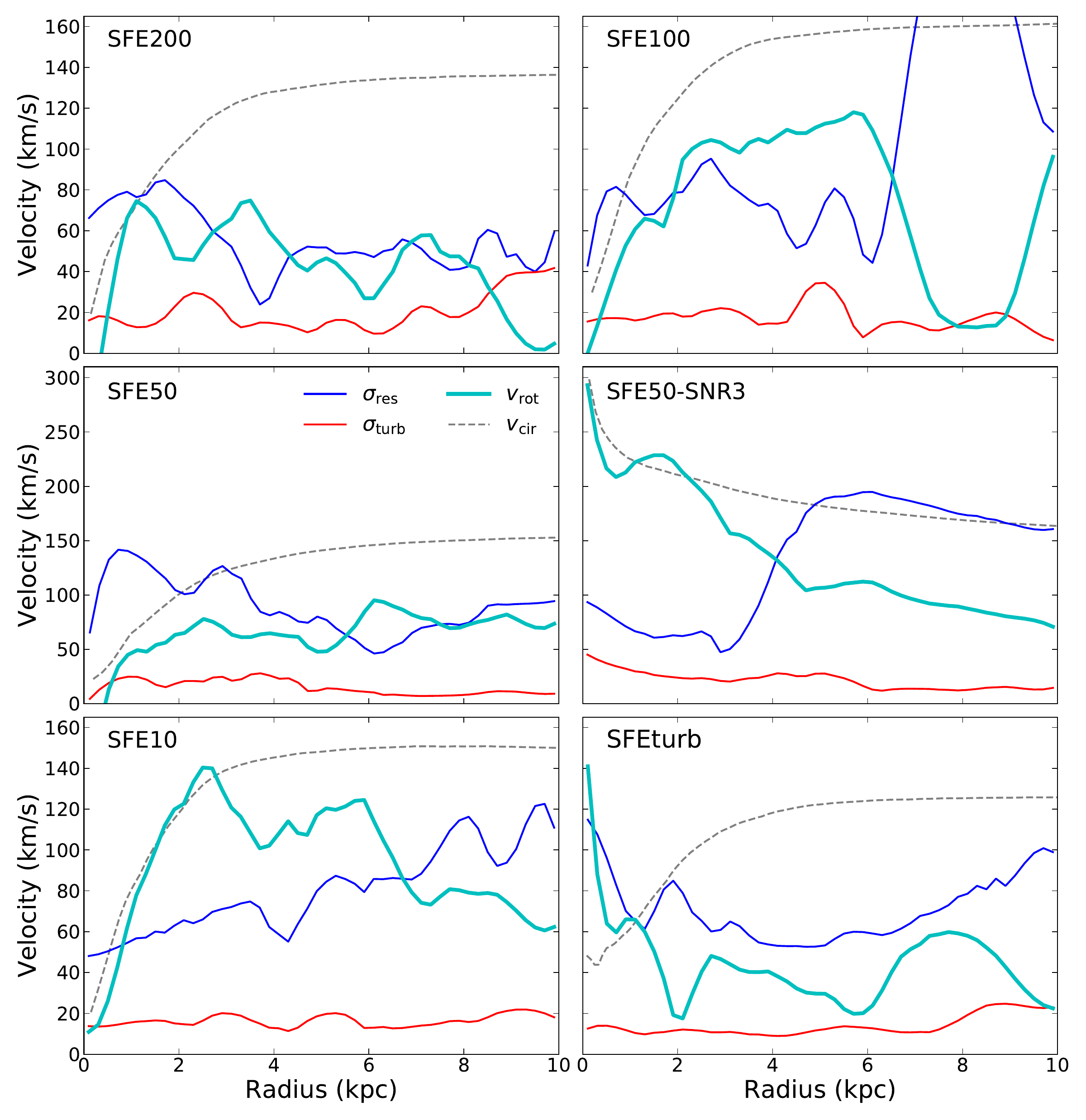}
    \vspace{-1mm}
\caption{Kinematics of the molecular gas in the simulated galaxies, calculated in cylindrical shells: rotation velocity (cyan), subgrid-scale turbulence (red), resolved turbulence (blue). Grey dashed lines show the circular velocity profile.}
    \label{fig:profile_veldis_gas}
\end{figure*}

Figure~\ref{fig:profile_radial_sumstar} compares the stellar profiles of galaxies in different runs. Even though our \revi{galaxies} are thick, all of them can still be fit reasonably well by an exponential \rev{disc} model with the scale length $h_R = 1.1 \pm 0.1$~kpc, in the radial range $2-6$~kpc. The only exception is run SFE50 that has $h_R\approx 1.5$~kpc. Since the gas density profiles are shallower, formation of new stars over time would gradually increase the stellar scale length. \revi{We note that the exponential fits here exclude the central parts of the galaxies, which may harbor dense bulges. }

Interestingly, observations of clumpy galaxies at $z\simeq 1-2$ \citep[e.g.,][]{Shibuya:2016aa} find surface brightness profiles with the S\'ersic index of $n\simeq 1$, corresponding to an exponential \rev{disc}. For example, massive ($\Ms\sim 10^{9.6}\Msun$) star-forming galaxies at $1<z<3$ in the ZFOURGE survey can be fit by $n\approx 1.2$ and an effective radius of $R_e\approx 2$~kpc \citep{Papovich:2015aa,Forrest:2018aa}. This is very close to the effective radii we obtain for our galaxies: $R_e\approx 1.68\, h_R \approx 1.8$~kpc. Thus the stellar density profiles of our simulated galaxies are typical of star-forming galaxies at these intermediate redshifts.
\\
\\

\subsection{Velocity dispersion profiles}

The kinematics of the interstellar medium in the simulated galaxies is shaped to a large extent by turbulent flows. The ART code models the injection of unresolved (subgrid scale) turbulence by SN explosions, as well as turbulent cascade from resolved scales. Even though these turbulent motions decay on a dynamical timescale, they still play an important role in distributing the SN energy over the neighbouring cells in actively star-forming galaxies.

Figure~\ref{fig:profile_veldis_gas} shows kinematic profiles of the molecular gas in all six galaxies. Using the previously defined centre and orientation of the \rev{galaxy plane}, the rotation curve is calculated in linearly spaced cylindrical shells of 200 pc in width. The SGS turbulence is averaged over the cells falling into the cylindrical shell. \revi{The resolved velocity dispersion $\sigma_{\rm res}$} is the residual motion of cells, after subtracting the mean rotation velocity. Both dispersions are calculated in three dimensions, weighted by $\Hm$ gas mass. When averaged over the cylindrical shells, the resolved dispersion dominates over the unresolved SGS turbulence. The resolved dispersion is comparable to the rotation velocity by amplitude and typically anticorrelates in radius.

Two runs stand out from the general trend. The weaker feedback run SFE50-SNR3 forms a massive bulge, which leads to the sharp rise of the circular velocity at the centre. The striking difference with the parallel run SFE50 with the same $\epsff$, where the boost factor is only 5/3 times larger, may seem surprising. In the inner few kpc the rotation velocities differ by more than a factor of two. A likely cause of such a difference is in the unstable nature of formation of dense stellar clumps. If a dense bulge happens to begin forming, as was the case in run SFE50-SNR3, then stellar feedback is insufficient to halt its growth until most of the inner gas is converted into stars. In the other case when momentum feedback is always sufficiently strong, massive clumps do not form and subsequent feedback succeeds in regulating star formation. The other run that shows a rising rotation velocity towards the centre is SFEturb. The last available output for this run is at a higher redshift than the rest and the magnitude of the velocity increase is relatively small. It is difficult to predict whether this increase will persist to later epochs or conform to the other runs.

The turbulent flows of $\Hm$ gas are highly supersonic. The 3D turbulent velocities reach $100-150$~km~s$^{-1}$ and violently stir the gas clouds in the \rev{galaxy plane}. These flows are responsible for the thick structure of the gaseous \rev{discs} of all our simulated galaxies. They also create significant vertical motion of the gas clouds.
\\
\\
\\

\section{Toomre analysis} \label{sec:toomre}

The local stability criterion, first derived by \citet{safronov60} and \citet{toomre64} and reviewed in \citet{toomre77}, has been commonly used to identify star-forming regions in regular \rev{disc} galaxies. For a gaseous \rev{disc}, the $Q$ parameter is defined as: 
\begin{equation}
   Q = \frac{\sigma \kappa}{\pi G\Sigma},
   \label{eq:Q}
\end{equation}
where $\kappa$ is the epicycle frequency, $\sigma$ is the velocity dispersion, and $\Sigma$ is the surface mass density. A self-gravitating \rev{disc} is stable to axisymmetric perturbations on scale $\lambda$ if
\begin{equation}
   Q > 2 \left( \frac{\lambda}{\lcrit} - \frac{\lambda^2}{\lcrit^2} \right)^{1/2},
   \label{eq:Q_lambda}
\end{equation}
where $\lcrit$ is the largest unstable wavelength for a zero-pressure ($\sigma=0$) \rev{disc}:
$$
   \lcrit \equiv \frac{4\pi^2 G \Sigma}{\kappa^2}.
$$ 
The \rev{disc} with $Q>1$ is stable on all perturbation scales $\lambda$, but even smaller $Q$ can indicate stability for $\lambda < \frac{1}{2}\lcrit$ \citep{binney_tremaine08}.

The wavelength of the fastest growing perturbation is not $\lcrit$ but instead is given by \citep[e.g.,][]{Nelson:2006aa}
\begin{equation}
   \lambda_T \equiv \frac{2 \sigma^2}{G \Sigma} = \lcrit \frac{Q^2}{2}.
   \label{eq:lambdaT}
\end{equation}
It is a two-dimensional analogue of the Jeans wavelength.

The amount of gas contained within a circle of diameter $\lambda_T$ is an expected mass that would collapse into a self-gravitating object, called the Toomre mass:
\begin{equation}
   M_T \equiv \frac{\pi}{4} \lambda_T^2 \Sigma = \frac{\pi \sigma^4}{G^2 \Sigma}.
   \label{eq:MT}
\end{equation} 
In some recent work \citep[e.g.,][]{reina-campos_kruijssen17, pfeffer_etal18}, the Toomre mass was defined alternatively as the mass within $\lcrit$:
\begin{equation}
   \tilde{M}_T \equiv \frac{\pi}{4} \lcrit^2 \Sigma = \frac{4 \pi^5 G^2 \Sigma^3}{\kappa^4}.
   \label{eq:MTtilde}
\end{equation} 
The two definitions are related by
$$
   M_T = \tilde{M}_T \, \frac{Q^4}{4}.
$$
For dense clumps with $Q<1$, the alternative $\tilde{M}_T$ can significantly overestimate the Toomre mass.

The original Toomre criterion applies to a fully gaseous\rev{, rotating, thin disc, under linear perturbations}. Real galaxies, and our simulated galaxies, contain gas, stars, and dark matter. While the dark matter has much larger velocity dispersion and does not significantly affect \rev{disc} stability, stars have velocity dispersion comparable to gas and contribute additional gravitational force. Different velocity dispersions, and the corresponding different scaleheights above the \rev{disc} plane, of the stellar and gaseous components mean that they affect the stability criterion in a more complicated way. \citet{Rafikov:2001aa} presents a detailed analysis of the stability of multi-component \rev{discs} and provides a modified criterion:
\begin{equation}
   \frac{1}{Q_R} = \max_k \left\{ \frac{1}{Q_*}\frac{2\left(1-e^{-q^2}I_0(q^2)\right)}{q} + \frac{1}{Q_g}\frac{2q\xi}{1+q^2{\xi}^2} \right\} 
   \label{eq:Q_R}
\end{equation}
where $q\equiv k{\sigma}_*/\kappa$, $\xi\equiv {\sigma}_g/{\sigma}_*$, $I_0$ is modified Bessel function of the first kind, and $Q_*$ and $Q_g$ are the separate parameters for stars and gas. The maximum is taken over values of the perturbation wavenumber $k=2\pi/\lambda$. This criterion has been adopted in some simulations of galactic \rev{discs} \citep[e.g.,][]{Li:2005aa}.

\citet{Romeo:2011aa} and \citet{Romeo:2013aa} suggest a simplified version of the modified $Q$ parameter and show that it reasonably accurately reproduces Rafikov's criterion. The modified parameter for $N$ components of the matter distribution is
\begin{equation}
  Q_N = \left( \sum_{i=1}^N \, Q_i^{-1} \, \frac{2\sigma_m \sigma_i}{\sigma_m^2 + \sigma_i^2} \right)^{-1}
  \label{eq:Qeff}
\end{equation}
where $Q_i$ are the separate parameters for each component:
\begin{equation}
  Q_i \equiv \frac{\sigma_i \kappa}{\pi G\Sigma_i}
\end{equation}
and $\sigma_m$ is the velocity dispersion of the component with the lowest $Q_i$. The ratio of dispersions in equation~(\ref{eq:Qeff}) acts as a weighting factor that reduces the contribution of the other components with $\sigma_i \ne \sigma_m$. They also suggest additional correction to account for the thickness of different components. This correction is small and we do not use it in our analysis for clarity, but we denote their full formalism for the Toomre parameter as $Q_{RW}$.

\citet{Romeo:2013aa} show that their definition of $Q_N$ with $N=2$ can accurately approximate Rafikov's criterion. The definition of $Q_2$ was adopted, for example, in the model of \citet{Krumholz:2018aa}. To examine this approximation we calculated and compared the values of $Q_R$ and $Q_2$ for our galaxies, and found that indeed they agree to better than 5\%. The reason for such a close match can be understood as follows. If $Q_g < Q_*$, then the maximum of equation (\ref{eq:Q_R}) is reached when $q\xi = 1$, and the weighting factors before $Q_g$ are unity for both expressions ($Q_R$ and $Q_2$) while the weighting factors before $Q_*$ evaluate to be within 6\% of each other. If instead $Q_g > Q_*$, then the maximum of equation (\ref{eq:Q_R}) is reached when $q = 1$, and the weighting factors before $Q_g$ are the same while the weighting factors before $Q_*$ are within 7\% of each other. Thus in general the two expressions $Q_R$ and $Q_2$ differ by no more than a few percent.

For the calculation of $Q$ in our simulations we use equation~(\ref{eq:Qeff}) and consider three components: molecular gas ($\Hm$ multiplied by 1.32 to account for helium within molecular clouds), atomic gas (HI plus HeI), and stars.

The wavelengths $\lcrit$ and $\lambda_T$ can be evaluated using the modified surface density:
\begin{equation} \label{eq9}
  \Sigma_{\rm eff} \equiv \sum_{i=1}^N \, \Sigma_i \, \frac{2 \sigma_m^2}{\sigma_m^2+\sigma_i ^2}
\end{equation}
as
\begin{equation} \label{eq10}
  \lcrit = \frac{4\pi^2 G \Sigma_{\rm eff}}{\kappa^2}, \quad 
  \lambda_T = \frac{2 \sigma_m^2}{G \Sigma_{\rm eff}}.
\end{equation}

If the velocity dispersions of any two components are equal, their surface densities can be effectively combined. For example, if the dispersions of atomic and molecular gas are the same, $\sigma_{\HI} = \sigma_{\Hm} = \sigma_g$, then their $Q_i$ parameters can be combined into one 
$$
   Q_g = (Q_{\HI}^{-1} + Q_{\Hm}^{-1})^{-1} = \frac{\sigma_g \kappa}{\pi G (\Sigma_{\HI} + \Sigma_{\Hm})}.
$$
Analogously, if stars have the same velocity dispersion as the gas, their surface density can be added to the sum in the denominator. For example, \citet{orr_etal18} included the sum of the gas and stars to obtain the total mass surface density, $\Sigma = \Sigma_{\rm g} + \Sigma_*$, in their analysis of the FIRE simulations. 

Using the simulation outputs at the last available epoch, we calculate projected maps of $Q$-values in square patches of physical length $L=0.2\,$kpc. The orientation of the patches is aligned with the \rev{galaxy} plane, and the projection is along the direction perpendicular to the \rev{galaxy plane}.

The size $L$ is chosen to contain many gas cells for sufficient averaging, such that numerical discreteness does not affect the conclusions. The size of individual cells containing most of the neutral gas varies from about 50 to $100\,$pc at redshift $z=1.5$. \rev{The mass-weighted size of individual cells at the redshift z=1.5 outputs varies from 36 to 54 pc when weighted by molecular gas mass, or from 92 to 112 pc when weighted by neutral gas mass.} These cells are not at the highest level of refinement but dominate the gas mass. Therefore, we restrict the square patches to be no smaller than $200\,$pc. 

Below we describe the calculation of the three variables determining the $Q$ parameter: epicycle frequency, surface density, and velocity dispersion.

\subsection{Epicycle frequency}
\label{sec:kappa}

For an axisymmetric \rev{disc} with near-circular orbits \citep[e.g.,][]{binney_tremaine08}
\begin{equation}
  \kappa^2 = \kappa_R^2 \equiv \frac{2\bar{V}^2}{R^2} \left(1 + \frac{d\ln{\bar{V}}}{d\ln{R}}\right)
  \label{eq:kappaR}
\end{equation}
where $\bar{V}$ is the \rev{average velocity of circular motion}. We approximate it by a spherically-symmetric expression for the circular velocity:
$$
   V_{\rm c}(R) \equiv \left( \frac{GM(R)}{R} \right)^{1/2}.
$$
The actual circular velocity for a razor-thin exponential \rev{disc} is given by the Bessel functions, but the difference is only $\sim 10\%$, which is below other approximations necessary for our analysis.

In the limit of a flat rotation curve, \rev{the expression simplifies to}
\begin{equation}
   \kappa^2 = \kappa_c^2 \equiv \frac{2V_{\rm c}^2}{R^2}.
   \label{eq:kappac}
\end{equation}
We define separate variables $\kappa_R$ and $\kappa_c$ to refer to these commonly used approximations. \rev{Below we use equation~(\ref{eq:kappac}) as the definition of $\kappa_c$ and use it to evaluate $\kappa_c$ even when $V_{\rm c}$ is not constant.}

Our simulated galaxies have much smaller and irregular rotation pattern because of additional pressure support provided by turbulent motions, as shown in Figure~\ref{fig:profile_veldis_gas}, and therefore we do not expect these approximations to hold. Since the epicycle expansion is done around the mean azimuthal velocity, a more appropriate quantity is the actual rotation velocity $\bar{V}=V_{\rm rot}$. To verify the accuracy of calculation of $\kappa$, we also evaluated the expression 
\begin{equation}
  \kappa^2 = \kappa_\phi^2 \equiv \frac{2V_{\rm rot}^2}{R^2} \left(1 + \frac{d\ln{V_{\rm rot}}}{d\ln{R}}\right).
   \label{eq:kappaphi}
\end{equation}

\begin{table*}
\centering
\caption{Accuracy of calculation of $\kappa$ in various approximations} \label{tab:kapparatio}
\begin{tabular}{lccc}
      \toprule
     & $\kappa_c/\kappa_R$ & $\kappa_\phi/\kappa_R$ & $\kappa_T/\kappa_R$\\ 
     Run & 25-50-75\% range & 25-50-75\% range & 25-50-75\% range\\
      \midrule
 SFE200            & 0.79 - 0.84 - 0.94 & 0.46 - 0.57 - 0.76 & 0.60 - 0.86 - 1.17 \\ 
 SFE100            & 0.78 - 0.82 - 0.92 & 0.57 - 0.66 - 0.79 & 0.61 - 0.84 - 1.13 \\
 SFE50             & 0.74 - 0.81 - 0.96 & 0.42 - 0.52 - 0.73 & 0.61 - 0.89 - 1.21 \\
 SFE10             & 0.75 - 0.86 - 0.96 & 0.61 - 0.89 - 1.07 & 0.60 - 0.90 - 1.26 \\
 SFEturb           & 0.75 - 0.85 - 1.03 & 0.27 - 0.62 - 1.14 & 0.58 - 0.86 - 1.19 \\
 SFE50-SNR3        & 1.03 - 1.09 - 1.13 & 0.58 - 0.74 - 0.96 & 0.49 - 0.76 - 1.11 \\
      \bottomrule
\end{tabular}
\end{table*}

\begin{figure*}
  \centering
  \includegraphics[width=\textwidth]{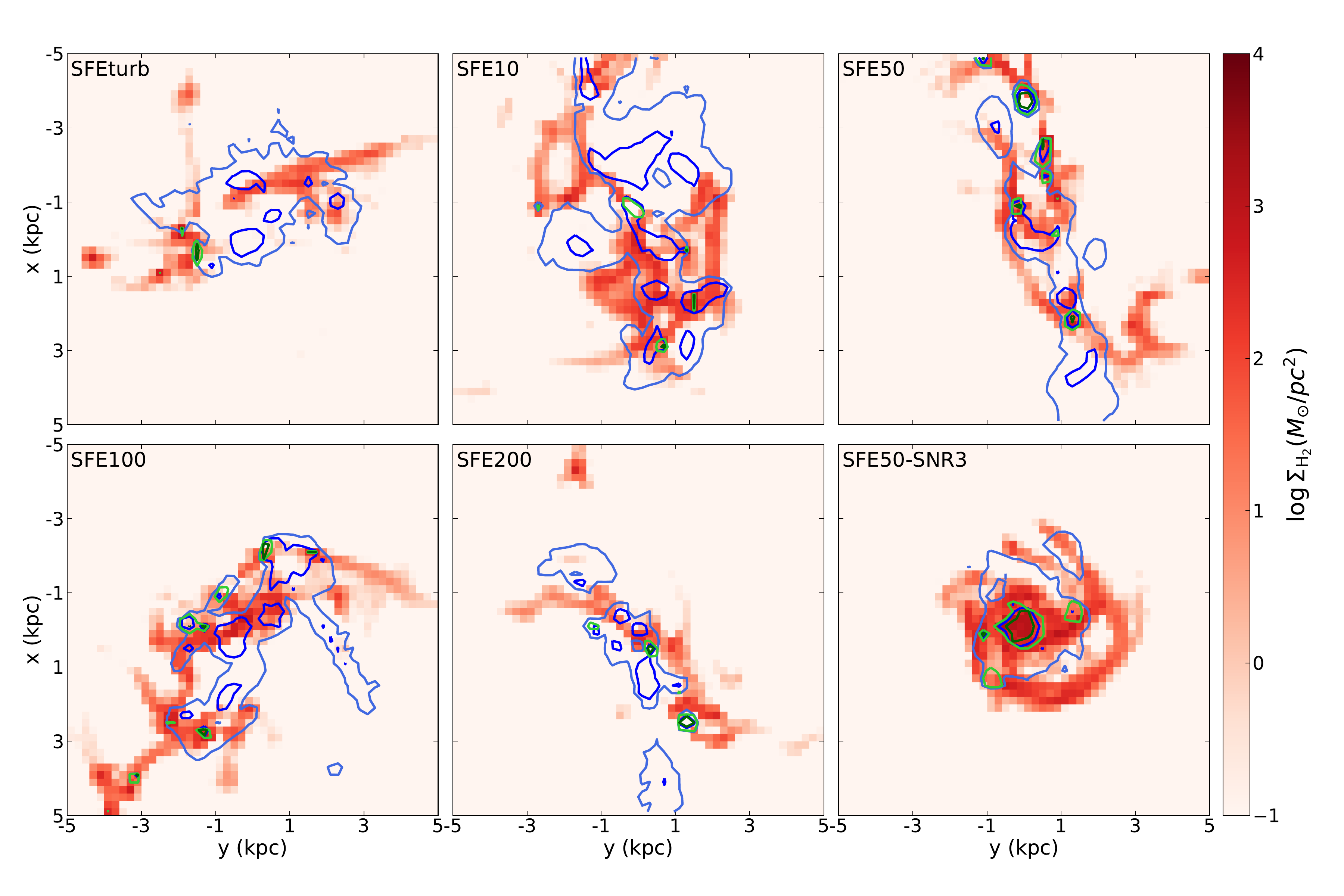}
  \vspace{-6mm}
\caption{Map of $\Hm$ surface density with projection thickness of $\pm 2\,$kpc. Blue contours enclose 80\% and 99\% of mass of young stars formed in last 50~Myr. Green contours enclose the same mass fractions of young stars formed only in last 10~Myr.} 
  \label{fig:H2map}
\end{figure*}

Table~\ref{tab:kapparatio} shows the distributions of ratios of $\kappa$ given by equations~(\ref{eq:kappaR}-\ref{eq:kappaphi}). The epicycle frequencies are calculated in axisymmetric cylindrical shells of width $L=200\,$pc. As expected, the median $\kappa_c$ is consistently lower than $\kappa_R$ by $\sim 15\%$ because the circular velocity curves are still increasing in the range of radii we consider, $R < 5\,$kpc. However, the difference is small.

\begin{figure*}
\centering
\includegraphics[width=1.0\textwidth]{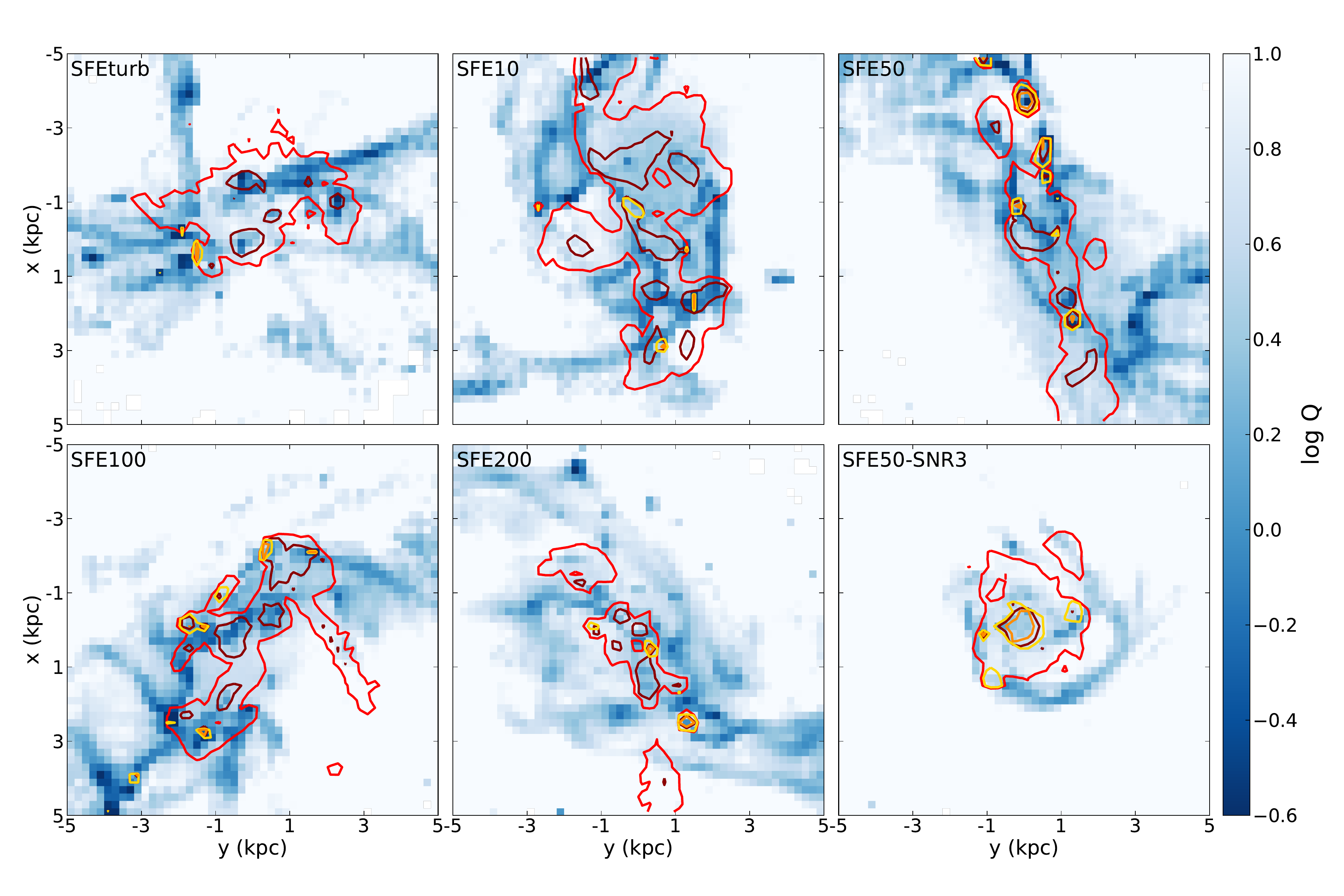}
\vspace{-6mm}
\caption{Map of the $Q$ parameter in all six runs, at the last available output. Contours in red show the surface density of stars younger than 50~Myr enclosing 80\% and 99\% of the mass of these stars. Yellow contours enclose the same mass fractions of young stars formed only in last 10 Myr. The values of $Q$ are calculated using the $\kappa_c$ definition to avoid visual artifacts.} 
  \label{fig:Q_Qgas_all}
\vspace{6mm}
\end{figure*}

The values calculated directly from the rotation velocity have a larger spread from $\kappa_R$, by $\sim 30\%$ in the median. The interquartile range of the distribution of $\kappa_\phi/\kappa_R$ is also wider, extending from 0.3 to 1.1. However, despite this significant scatter from shell to shell, the overall good accuracy of calculation of $\kappa$ using the simple approximation of equation~(\ref{eq:kappaR}) is surprising. It validates our analysis of the $Q$ parameter to better than a factor of two. For the distributions of $Q$ parameter described below the epicycle frequency is calculated in cylindrical shells within the \rev{galaxy} plane using equation~(\ref{eq:kappaR}), except in maps shown in Figure~\ref{fig:Q_Qgas_all} where we use equation~(\ref{eq:kappac}) to avoid visual artifacts.

\citet{Inoue:2016aa} used their simulations of high-redshift galaxies to compare the calculation of $\kappa$ using the circular velocity $V_c$ and the actual rotation velocity $V_{\rm rot}$. In agreement with our results, they find that the latter gives somewhat lower values, but the difference is small. They also note that the rotation velocity can decrease with radius suddenly and lead to $\kappa^2 < 0$ in some shells in the outer parts of the galaxies.

In addition to the above approximations, we evaluate another expression for the epicycle frequency suggested by \citet{pfeffer_etal18}:
\begin{equation}
   \kappa^2 = \kappa_T^2 \equiv -\sum \lambda_i - \lambda_1, 
\end{equation}
where $\lambda_i$ are the eigenvalues of the tidal tensor around the patch. The maximum eigenvalue is $\lambda_1 \ge \lambda_2, \lambda_3$. We calculate the tidal tensor on the uniform grid of 200~pc patches and evaluate the ratio of $\kappa_T/\kappa_R$ on these patches, instead of the cylindrical shells as for the other axisymmetric definitions of $\kappa$. The ratio $\kappa_T/\kappa_R$ is similar to $\kappa_c/\kappa_R$ but with larger scatter. The absolute values of ${\kappa}_R$ are similar to those in Fig.~A1 of \citet{pfeffer_etal18}: ${\kappa}_R$ decreases from around $200\,{\rm Gyr^{-1}}$ in inner 1 kpc to $\sim 30\,{\rm Gyr^{-1}}$ at $R=5$ kpc, for all runs except SFE50-SNR3.

\subsection{Surface density}

We calculate the surface densities of all components by projecting over a column of thickness $\pm 2\,$kpc around the \rev{galaxy} plane. This thickness was chosen to contain most (about 80\%) of the gas and stars. Only in SFE200 run such a column contains 70\% of the mass. To account for 80\% of the mass in that run we would need to integrate within $\pm 3\,$kpc, but this difference is not essential to our analysis and we chose to keep the column thickness the same for all runs.

In most patches the gas dominates the surface density. The median ratio $\Sigma_*/\Sigma_g$ is in the range $0.7-0.9$ for most runs. The two exceptions are SFE10 with $\Sigma_*/\Sigma_g \approx 1$ and SFE50-SNR3 with $\Sigma_*/\Sigma_g \approx 2$. 
Most of the gas near the \rev{galaxy plane} is atomic and some is ionized. Below we investigate specifically the distribution of molecular $\Hm$ gas, which is directly linked to star formation.

Figure~\ref{fig:H2map} compares the maps of molecular gas density with star formation rate density. To calculate the star formation rate in a large patch of a galaxy within a given interval of time, we need to take into account a finite duration of formation of cluster particles in our CCF algorithm \citep{li_etal18}. At the very beginning of the formation episode of each particle, the formation rate is low. Then as particle mass rises, the gas accretion rate increases and the star formation rate picks up, until stellar feedback halts the accretion. The variable $\tau_{\rm ave}$ approximately corresponds to the peak time of formation rate. Therefore, we take the star particle "zero age" to be the moment when it has gone through one $\tau_{\rm ave}$ after the creation. The typical values are $\tau_{\rm ave} \sim 1-2$~Myr.

Surprisingly, Figure~\ref{fig:H2map} shows that for the majority of patches the regions of high $\Hm$ density \textit{do not} coincide with the regions of high SFR density. It is because strong stellar feedback quickly heats and removes the gas from star forming regions. This becomes an important point in the calculation of the depletion time in Section~\ref{sec:depletion}.

Observations of high-redshift galaxies \citep[e.g.,][]{Daddi:2010aa, Tacconi:2013aa, Dessauges-Zavadsky:2017aa} provide an analogous comparison of CO and radio continuum maps with optical, UV, and IR images. Observed emission from molecular gas and young stars is generally in the same place, but they do not coincide exactly. Other simulation results also show decoupled gas and SFR maps. For example, in the FIRE simulations, which use $\epsff=100\%$, \citet{Oklopcic:2017aa} find that gas clumps coincide with instantaneous SFR maps fairly well, but start to decouple from the SFR averaged over 10 Myr. The map of SFR averaged over 100 Myr shows that gas clumps do not trace the SFR peaks at all. This is similar to our results, where SF averaged over 10 Myr is located near the peaks of molecular gas, while SF averaged over 50 Myr correlates less well with the $\Hm$ map. Most of the molecular gas is not participating in star formation at any given time.

\begin{figure*}
\centering
\includegraphics[width=\textwidth]{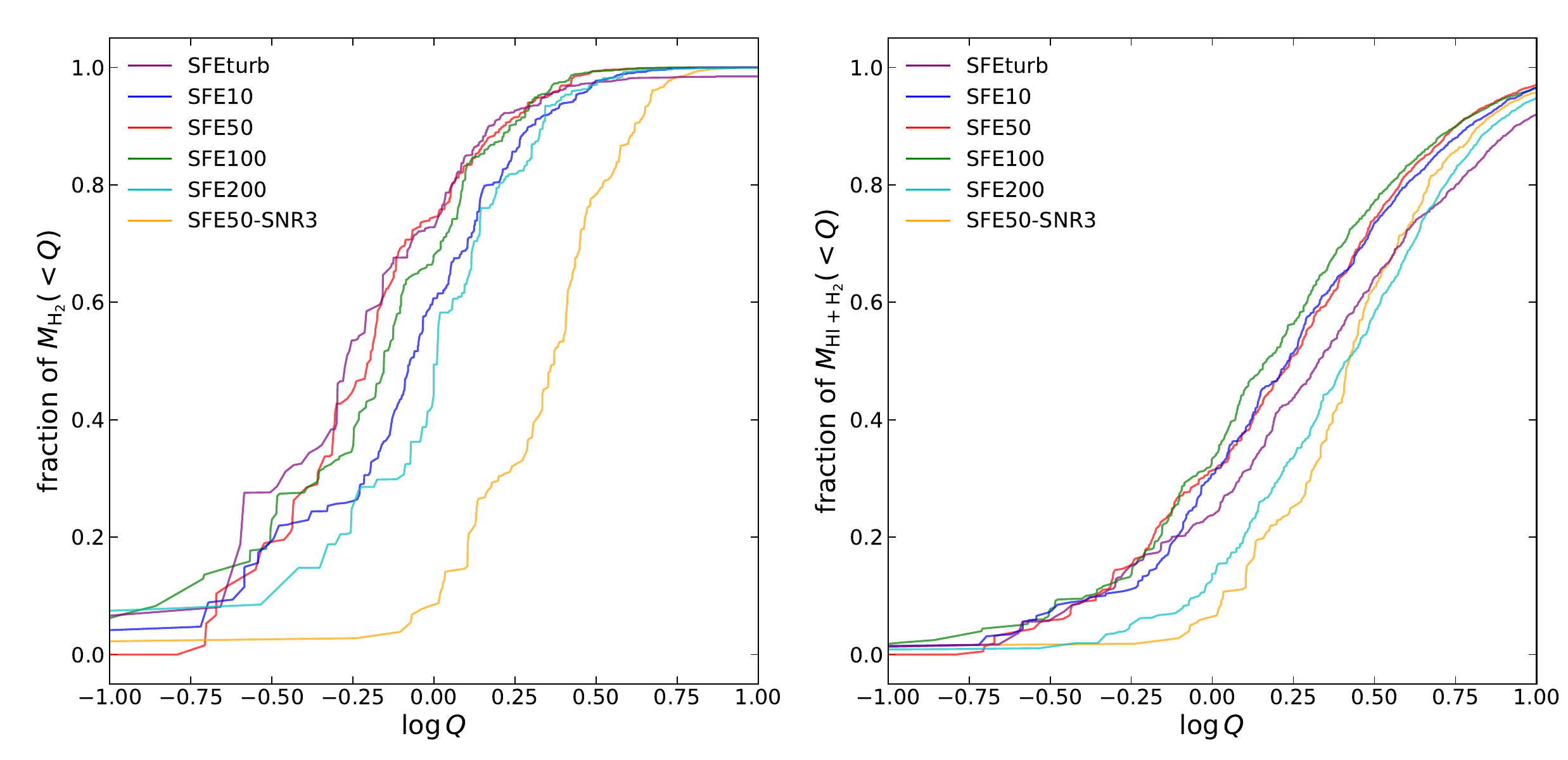}
\vspace{-5mm}
\caption{Cumulative distribution of the molecular $\Hm$ (left) and neutral $\Hn$ (right) gas mass within $\pm2\,$kpc as a function of $Q$. The velocity dispersion used to calculate $Q$ for the right panel is weighted by the $\Hn$ mass for consistency. The epicyclic frequency is calculated using the $\kappa_R$ definition.}
  \label{fig:sum_Q_all_cuH2}
\end{figure*}

\subsection{Velocity dispersion}

\rev{Since our galaxies are not thin, rotating discs with only velocity dispersion in radial direction, as in the original Toomre analysis, it is more reasonable to include all components of velocity dispersions as all of them resist gravitational collapse.} The full velocity dispersion of gas includes three components: the sound speed within a cell $c_s$, the subgrid scale turbulence (SGS; unresolved velocity dispersion) within a cell $\sigma_T$, and velocity differences between neighbour cells (resolved velocity dispersion) $\sigmacell$:
$$
   \sigma_g^2 = c_s^2 + \sigma_T^2 + \sigmacell^2.
$$
Using gas mass $M_k$ in a cell $k$ (see two versions below), we calculate the mass-weighted average of each component for the cells within a given patch:
\begin{eqnarray}
   c_s^2 & = & \frac{\sum_{k} M_k \, c_{s,k}^2}{\sum_{k} M_k} \nonumber\\
   \sigma_T^2 & = & \frac{2\sum_k E_{\mathrm{turb},k}}{\sum_k M_k} \label{eq:sigma_parts}\\
   \sigmacell^2 & = & \sum_{i=r,\phi,z}\left[\frac{\sum_k {v_{i,k}}^2 \,M_k}{\sum_k M_k} - \left(\frac{\sum_k v_{i,k} \, M_k}{\sum_k M_k}\right)^2\right].\nonumber
\end{eqnarray}    
Here $\mathbf{v}_{k}$ is the velocity of cell $k$ in the galacto-centric reference frame, and $E_{\mathrm{turb},k}$ is the energy in subgrid-scale turbulence in cell $k$, which is directly calculated in the simulation.

The goal of Toomre analysis is to identify regions of the ISM that are unstable to collapse and star formation. Which parts of the ISM are directly observable depends on the detection method. Radio observations of CO and HCN transitions probe dense molecular gas, while 21-cm line technique detects atomic hydrogen gas. To facilitate comparison with both types of observations, we consider two versions of the $Q$ parameter, which differ only in the weighting of the velocities. In one version the sound speed and the resolved velocity dispersion are weighted by the mass of molecular gas, $M_k = M_{\Hm}$, while in the other they are weighted by the mass of total neutral gas, $M_k = M_{\Hn}$. In the expression for the SGS turbulence, $M_k$ is always the total gas mass. For most of our analysis we use the result of weighting by the $\Hm$ mass; it is therefore implied by default. Only when we discuss the distribution of neutral gas, we use the $\Hn$ mass weighting for consistency.

We take the sum over all cells in a given square patch, projected over $\pm 0.8\,$kpc. We chose this smaller projection length than for the surface density to capture most of the molecular gas while avoiding the contamination by ionized gas that could skew the measurement of the sound speed. This choice also limits the resolved velocity dispersion. The column $\pm 0.8\,$kpc contains over 90\% of $\Hm$ mass in all runs except SFEturb (which contains 80\%).

Within patches of $L=0.2\,$kpc the SGS turbulence dispersion is typically larger than the resolved dispersion. The median ratio $\sigmacell/\sigma_T \approx 0.6$, when $\sigmacell$ is $\Hm$ mass weighted.
This situation is different from the azimuthal averages shown in Figure~\ref{fig:profile_veldis_gas}, where $\sigma_{\rm res}$ is the tangential velocity dispersion of cells in a cylindrical shell. Here instead we compute $\sigmacell$ using the actual local mean velocity in a patch. This reduces the residual dispersion substantially.

The velocity dispersion of stars is calculated simply as inter-particle dispersion, analogously to $\sigmacell$ but weighted by stellar mass. In most patches (over 75\%), the stellar dispersion dominates over the gas dispersion. The median ratio $\sigma_g/\sigma_* \approx 0.4$ for the molecular $\Hm$ gas and 0.5 for the neutral $\Hn$ gas.

\subsection{Toomre mass}
\rev{Using the epicycle frequency determined in Section~\ref{sec:kappa} and the effective $\lcrit$ and $\lambda_T$ (Equation~\ref{eq10}), we can now calculate the corresponding Toomre mass under the multi-component $Q$ definition.} We use patches of $200\,$pc, taking $\Sigma$ to be the gas surface densitiy projected over $\pm\,2$ kpc. Then we average $M_T$ and $\tilde{M}_T$ (equations~\ref{eq:MT}--\ref{eq:MTtilde}) in linearly-spaced cylindrical shells. Our values of $M_T$ generally increase with radius and vary from $10^{8.2}$ to $10^{10.6}\Msun$ (interquartile range) in the patches with $M_{\Hm}>10^6\Msun$ (see Section~\ref{sec:Qmaps} for justification of this threshold). However, $\tilde{M}_T$ is systematically higher; it ranges from $10^{8.6}$ to $10^{11}\Msun$ for the same patches. This is several orders of magnitude larger than the values found by \citet{pfeffer_etal18}, mainly because the gas surface density in our simulations is about an order of magnitude larger.

Note that \citet{Tamburello:2015aa} account for non-linear growth of perturbations in collapsing clumps and find that the actual fragmentation mass is lower than $M_T$ by a factor of several. Subsequent dynamics of collapsed clumps may also affect their mass because of merging and agglomeration with other clumps. We do not investigate the distribution of clump masses in this paper but plan to do it in follow-up work.

\subsection{Distribution of the \textit{Q} parameter}
\label{sec:Qmaps}

Figure~\ref{fig:Q_Qgas_all} shows projected maps of the $Q$ parameter and compares them with the star formation rate density. The maps of $Q$ generally follow the maps of gas density but pick out sharper features, such as spiral arms or filaments reaching to the galactic centre. The dynamic range of $Q$ is significantly reduced relative to that of the molecular gas surface density, which makes $Q$ a useful predictor of future star-forming regions.

An example of the weak feedback run (bottom right panel) illustrates that some high-density regions may not be unstable because of a steep potential well and high velocity dispersion. Thus a simple density threshold would not correctly pick gas patches that are unstable to gravitational collapse. 

In Figure~\ref{fig:sum_Q_all_cuH2} we show the cumulative distribution of $Q$, weighted by the mass of molecular (left panel) and all neutral hydrogen (right panel). That is, the left panel shows the fraction
$$
    \frac{M_\Hm(<Q)}{M_\Hm},
$$
and analogously for $\Hn$ in the right panel. The weaker feedback run SFE50-SNR3 shows a very different distribution, but all other runs with the same feedback strength show consistent results. The cumulative $\Hm$ masses are rising sharply around the median values $Q\approx 0.5-1.0$. These medians are remarkably close to unity, given the many approximations in our calculation of the $Q$ parameter and the irregular structure of these high-redshift galaxies. It also suggests that majority of the molecular gas is in the marginally stable dynamical state, which may indicate self-regulation of star formation by stellar feedback.

The distribution of neutral gas mass is shifted systematically towards higher values of $Q\approx 1.5-2.6$. The difference is mainly because the $\Hm$ weighting selects regions of higher surface density and slightly lower velocity dispersion, both of which reduce $Q$.

\rev{Table~\ref{tab:Qiqr} shows the interquartile ranges of the cumulative distribution of $Q$ in patches of $L=0.2\,$kpc, weighted by $\Hm$ surface density and $\Hn$ surface density, respectively.}

The Toomre analysis indicates that there is a threshold at $Q\lesssim 1$ to distinguish the unstable regions of the \rev{disc}. It is given by equation~(\ref{eq:Q_lambda}) that depends on the ratio of the perturbation scale to the largest unstable wavelength. To test the applicability of this analysis to the simulated galaxies, we calculated the largest unstable wavelength $\lcrit$ on the patches that are capable of forming stellar particles. Because of the minimum adopted mass of stellar particles in the simulations ($\sim 10^3\Msun$), patches with insufficient $\Hm$ mass are unable to produce even a single particle. This is a numerical resolution limitation and therefore, such patches should not be included in our analysis. In fact, the patch size is much larger than the size of star-forming regions adopted in our runs ($\sim 5\,$pc) and so the limiting mass should be significantly larger than the minimum particle mass. Accounting also for the low star formation efficiency in some of the simulations, we set the threshold $\Hm$ mass at $10^6\Msun$. Experimentation with lower values ($10^4-10^5\Msun$) showed that the median $\lcrit$ decreases as the threshold decreases, by $\approx 1-2$~kpc.
However, we choose the larger threshold value because it leads to a more reliable calculation of the depletion time in the next section.
 
\begin{table}
  \centering
  \caption{Distribution of $\lcrit$ and $\lambda_T$ in patches of $L=0.2\,$kpc} \label{tab:lambda_crit}
  \begin{tabular}{lcc}
   \toprule
       & $\lcrit$ (kpc) & $\lambda_T$ (kpc) \\
   Run & 25-50-75\% range & 25-50-75\% range \\
   \midrule
   SFE200            & 2.3 - 5.1 - 11.8 & 2.1 - 3.4 - \phantom{1}6.7 \\
   SFE100            & 3.4 - 6.4 - 16.0 & 1.5 - 2.5 - \phantom{1}4.7 \\
   SFE50             & 4.8 - 7.6 - 15.6 & 1.5 - 2.9 - \phantom{1}7.2 \\
   SFE10             & 2.8 - 4.2 - \phantom{1}6.2 & 1.4 - 2.2 - \phantom{1}5.1 \\
   SFEturb           & 3.1 - 4.8 - 10.2 & 0.9 - 1.3 - \phantom{1}2.6 \\
   SFE50-SNR3        & 1.1 - 1.6 - \phantom{1}2.5 & 3.0 - 5.9 - 11.7 \\      
   \bottomrule
  \end{tabular}
\end{table}

Table~\ref{tab:lambda_crit} shows the cumulative distributions of $\lcrit$ for patches of 0.2~kpc on the side with $\Hm$ mass above $10^6\Msun$. The critical wavelength is quite large, between 4 and 8 kpc in the median for all stronger feedback runs. Such large values are caused mainly by the high surface density of gas and stars in these high-redshift galaxies. 

The values of the Toomre wavelength $\lambda_T$ are systematically lower and range between $1-3$~kpc in the median. SFE50-SNR3 run is again an exception because of its higher $Q$ values. \rev{It is straightforward to show, with $Q_N, \lcrit, \lambda_T$ written in terms of $\sigma_m$ and $\Sigma_{\rm eff}$ following Eq.~(\ref{eq9}) and (\ref{eq10}), that t}he relation between the two wavelengths, given by equation~(\ref{eq:lambdaT}), is still valid even for the multi-component definition of the $Q$ parameter.

\begin{figure*}
  \centering
  \includegraphics[width=\textwidth]{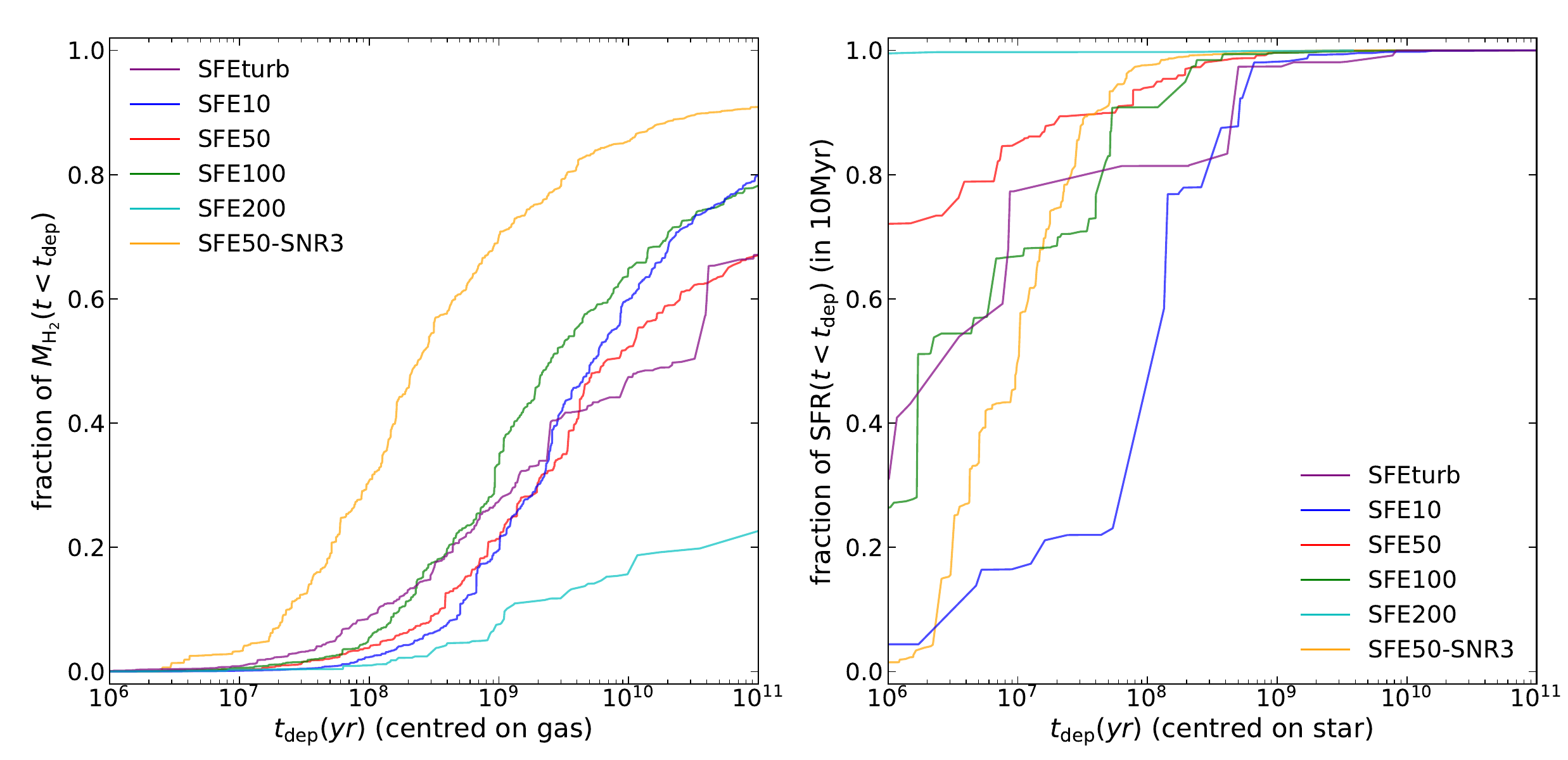}
  \vspace{-5mm}
  \caption{Cumulative distribution of the depletion time of molecular gas in patches of 100 pc in size. Left panel is for patches with $M_{\Hm}>10^6\Msun$ centered on gas density peaks, weighted by $\Hm$ mass. Right panel is for patches with non-zero SFR within 10~Myr centered on SFR peaks, weighted by SFR.}
  \label{fig:tdep_cu}
\end{figure*}

 \begin{table}
  \centering
  \caption{Distribution of $Q$ in patches of $L=0.2\,$kpc} \label{tab:Qiqr}
  \begin{tabular}{lcc}
   \toprule
     & $Q$ ($\Sigma_{\Hm}\,$weighted) & $Q$ ($\Sigma_{\Hn}\,$weighted)\\
   Run & 25-50-75\% range & 25-50-75\% range \\
   \midrule
   SFE200            & 0.56 - 1.02 - 1.39 & 1.38 - 2.55 - 4.46 \\
   SFE100            & 0.33 - 0.70 - 1.18 & 0.76 - 1.47 - 2.95 \\
   SFE50             & 0.37 - 0.63 - 1.03 & 0.76 - 1.73 - 3.25 \\
   SFE10             & 0.47 - 0.84 - 1.36 & 0.87 - 1.71 - 3.35 \\
   SFEturb           & 0.26 - 0.53 - 1.02 & 1.04 - 2.07 - 4.12 \\
   SFE50-SNR3        & 1.35 - 2.35 - 2.94 & 1.75 - 2.64 - 4.17 \\
   \bottomrule
  \end{tabular}
 \end{table}

The condition $Q<1$ could be used to select star-forming regions. The gas mass in patches that satisfy this condition is $\sim (1-5.5)\times 10^8\Msun$, which generally accounts for 60-70\% of the total $\Hm$ mass (except for the weaker feedback run, which has less than 10\% of the $\Hm$ mass in $Q<1$ patches). This mass is comparable to the mass converted to stars within 50~Myr. Except for SFE50-SNR3 run, the ratio of mass of young stars formed within 50 Myr to mass of $\Hm$ in patches with $Q<1$ varies between 0.34 and 3 for the different runs.
Therefore, we can expect that most of the present molecular gas would be converted to stars on a timescale 50-100~Myr. This estimate is very approximate because Figure~\ref{fig:Q_Qgas_all} shows that the star forming regions do not align with the unstable $Q<1$ patches after 50 Myr. The match is better for very young star formation within only 10 Myr. In the next section we investigate the gas consumption timescale in detail, and consider different patch sizes and different ways of centering the search region. 

\begin{figure*}
  \centering
   \includegraphics[width=\textwidth]{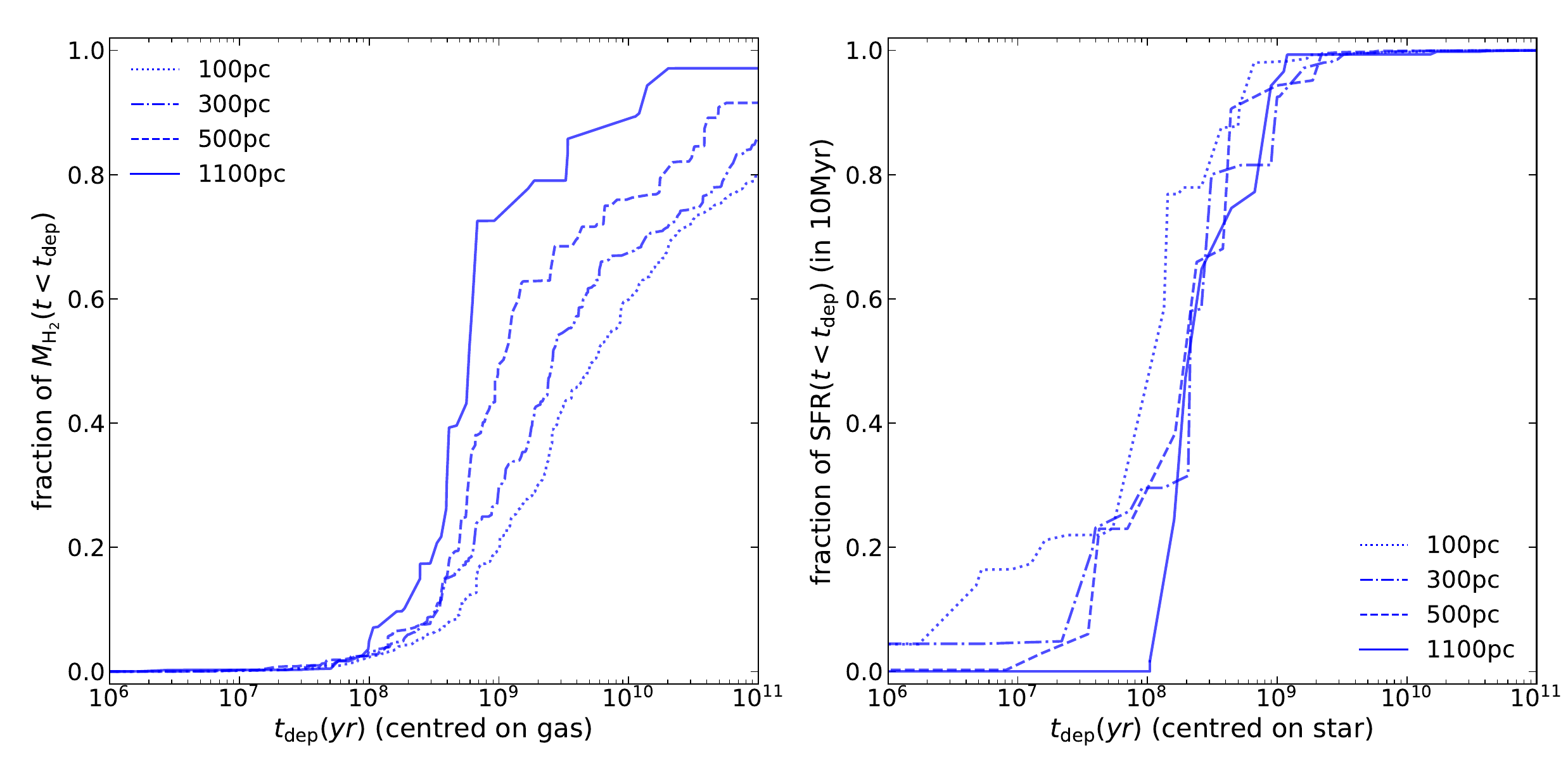}
   \vspace{-5mm}
\caption{Cumulative distribution of $\Hm$ depletion time for "centered-on-gas" (left panel) and "centered-on-stars" (right panel) versions for patches of different size. Here we show run SFE10 as an example. For "centered-on-gas" version, we use patches with $M_{\Hm}>10^6\Msun$ and weight them by $\Hm$ mass. For "centered-on-stars" version, we use patches with non-zero SFR and weight them by SFR. As patch size increases, the distribution shifts to smaller $\tdep$ for "centered-on-gas" case and to larger $\tdep$ for "centered-on-stars" case, similarly to the trend shown in Figure~\ref{fig:tdep}.} \label{fig:tdep_cu_scale}
\end{figure*}

\section{Gas depletion time} \label{sec:depletion}

An important measure of the global efficiency of star formation in galaxies is the gas depletion time. For our patches of size $L$, we define it as the ratio of the molecular gas surface density to the surface density of star formation:
\begin{equation}
  \tdep(L) = \frac{\Sigma_{\mathrm{H}_2}(L)}{\Sigma_\sfr(L)}.
\end{equation}

\citet{Schruba:2010aa}, \citet{Feldmann:2011aa}, and \citet{Kruijssen:2014aa} showed that the value of the depletion time depends on the spatial scale at which it is measured. It also depends on how the densities of gas and stars are calculated. In most observations they are counted in circular apertures centred on peaks of SFR density. Such regions may not already contain most of the gas they had before the onset of the star formation episode, resulting in relatively short depletion timescales. On the other hand, if the apertures are centred on peaks of the current gas density, the depletion times appear systematically longer.

To compare these two approaches, we use the following algorithm to calculate the surface densities of gas and SFR in patches of 100, 300, 500, and 1100~pc. First, we cover the plane of the galaxy with a rectangular grid of 100$\times$100 cells, each 100~pc wide and $\pm2$kpc thick, and calculate $\Sigma_{\mathrm{H}_2}$ and $\Sigma_\sfr$ in each cell. To go to larger size of 300~pc, we search for maximum of $\Hm$ mass (or SFR), record the position of that peak, group the peak and its surrounding $3^2-1=8$ cells into a larger patch and sum the $\Hm$ mass and SFR in this group. We label these 9 cells as "counted" and repeat the loop for the yet uncounted cells. Now we have a list of patches of 300~pc. Then we repeat this procedure for the chosen peaks and group $5^2$ cell to obtain patches of 500~pc, etc.

Using this algorithm, we calculate the depletion time $\tdep$ of $\Hm$ gas in patches of different size from 100~pc to 1.1~kpc. The patches are chosen to be centred on peaks of $\Hm$ mass or SFR. For the "centred-on-gas" version, we calculate the median of $\tdep$ in patches with $M_{\Hm}>10^6\Msun$ to eliminate the low-density regions that would be unable to form stellar particles in our cluster formation algorithm, and count the SFR averaged over 50~Myr for better statistics. For the "centred-on-SFR" or "centred-on-young-stars" version (which we call "centred-on-stars" for brevity), we consider patches with any non-zero SFR, and count the SFR averaged over 10~Myr in order to approximate more closely the instantaneous star formation.

To show the full range of $\tdep$ in both versions, in Figure~\ref{fig:tdep_cu} we plot the cumulative distribution of $\tdep$ for patches of the smallest size (100 pc). In the "centred-on-gas" version (left panel), the cumulative distributions of $\tdep$ weighted by $\Hm$ mass do not reach 100\%. This happens because some patches contain no stellar particles and therefore have formally infinite depletion time. The number of such patches is particularly large for SFE200 run, which has strongly misaligned molecular gas and young stars (see Figure~\ref{fig:H2map}). For the median $\Hm$ mass the depletion time ranges between $10^{9.5}$ and $10^{10.5}$~yr in the strong-feedback runs.

In contrast, the right panel shows that the depletion time in the "centred-on-stars" version is significantly shorter, typically below $10^{8}$~yr. Such discrepancy between the two counting methods is due to the gas-star misalignment which we emphasized above. Most of the star-forming sites have so little gas left within 100~pc that it would be exhausted in a relatively short interval of time.

Both versions of the gas depletion timescale vary strongly with the spatial scale on which they are calculated. To illustrate this dependence, we choose two representative runs (SFE50 and SFE10) and in Figure~\ref{fig:tdep_cu_scale} we plot cumulative distributions of $\tdep$ for different patch sizes. The distribution of $\tdep$ in the "centered-on-gas" version shifts to smaller values as patch size increases, by an order of magnitude between 100~pc and 700~pc. This shift is monotonic with the patch size and similar for the two runs shown.

The "centered-on-stars" version shows the opposite trend: $\tdep$ shifts to larger values. This change is roughly monotonic but differs between the two runs. In one case where $\tdep$ was very short on 100~pc scale (SFE50) the increase is dramatic, by about two orders of magnitude. In the other case with larger $\tdep$ (SFE10) the increase is only by a factor of two.

Adopting lower values of the threshold $M_{\Hm}>10^4-10^6\Msun$ increases the fraction of patches with infinite $\tdep$, leading to larger scatter in $\tdep$. If the threshold is taken to be $10^4\Msun$, even the median value of $\tdep$ in SFE200 run is infinite for all patch sizes.
In the other runs, the median value of $\tdep$ changes within 0.5~dex for the smallest patch size, and correspondingly less for larger patch sizes. On the other hand, increasing the threshold mass greatly reduces the available number of patches, so we do not set the threshold above $10^6\Msun$.

Despite the above variations, the estimate of $\tdep$ in both versions for all runs approaches a similar common range at the largest considered scale, $L=1.1$~kpc. This convergence is illustrated in Figure~\ref{fig:tdep}. The median values over patches for different runs are all contained within $10^8 - 10^9$~yr. Unlike Figures~\ref{fig:tdep_cu} or \ref{fig:tdep_cu_scale}, here the statistics of patches are not weighted by gas mass or SFR. The convergence is not strictly monotonic in all runs and there is large scatter from patch to patch. It is illustrated by shaded regions for one run; the amount of scatter is typical of all runs. In Section~\ref{sec:discussion_depletion} we compare our results with the expectation of models of galactic star formation and available observational estimates.

The depletion times for gas-centered and star-centered patches do not match exactly on the largest scale, because we average the SFR over different timescales in the two cases: 50~Myr for gas-centered patches and 10~Myr for star-centered patches. The reason for using different timescales is that young stars and molecular gas coincide little in our simulations due to strong feedback. Averaging the SFR in 50~Myr for star-centered $\tdep$ would lead to values smaller by several orders of magnitude.

We have also checked how the estimate of the depletion time varies with time, by examining previous outputs of each run. For the weaker feedback run SFE50-SNR3, the difference on all scales is small. For the other runs, the depletion time on kpc scale generally changes by factor of a few, while on smaller scales the difference is larger. The variation of $\tdep$ is larger for "center-on-stars" patches than for "center-on-gas" patches, because the value of SFR over 10~Myr is more stochastic, leading to larger scatter in $\tdep$. Galaxies going through a merger show more divergent estimates of $\tdep$ for both "center-on-gas" and "center-on-stars" versions on all scales, because of smaller overlap of SFR and molecular gas. The scatter of depletion time in previous outputs is similar to the scatter in snapshots shown in Figure~\ref{fig:tdep}. 

\subsection{Dependence of the depletion time on gas metallicity}

We calculated the median values of the $\Hm$ depletion time for patches of interest with metallicity higher and lower than the median metallicity. For almost all runs and all patch sizes, in both versions, the median $\tdep$ of patches with high metallicity is smaller than that of patches with low metallicity. However, the whole metallicity distribution is contained to a very narrow range (the interquartile range is smaller than 0.1$\,Z_{\odot}$ for all runs except SFE50-SNR3) such that the difference between the "high-metallicity" and "low-metallicity" values cannot be expected to lead to any substantial differences in physical properties of the gas. At the same time, the SFR density has a very large spread of several orders of magnitude in both cases. Therefore, we think that the dependence of $\tdep$ on metallicity cannot be robustly determined with our data.

\section{Discussion} \label{sec:discussion}

\subsection{Toomre analysis}

The Toomre $Q$ parameter has been measured in several observational studies, at high and low redshift. \citet{Genzel:2011aa} mapped four $z\approx2$ star-forming galaxies in gaseous $Q$, including a correction for multiple components. They find that ${\rm H}_{\alpha}$ clumps marking young stellar systems are present at the locations of gravitationally unstable gas ($Q\lesssim1$). However, this correspondence could result from their indirect inference of the gas density by using the same ${\rm H}_{\alpha}$ flux coupled with the global star formation relation. Follow-up work by \citet{Genzel:2014aa} find values of $Q<1$ in the outer regions of the observed galaxies and an increase towards the center, which they associate with higher central mass concentration and larger $\kappa$.
 
\citet{Romeo:2011aa} and \citet{Romeo:2013aa} used their $Q_2$ and $Q_3$ definitions to calculate the Toomre parameter for a sample of nearby spiral galaxies from the THINGS survey of \citet{Leroy:2008aa}. They find values of $Q\approx 2-5$, and no strong trend with galactocentric radius. In one third of the galaxies $Q$ is dominated by $\Hm$ in the inner parts and in the rest it is dominated by stars at all radii. \citet{Westfall:2014aa} used integral field spectroscopy for 27 nearby face-on spiral galaxies from \citet{Martinsson:2013aa} to calculate $Q_{RW}$, which is equivalent to $Q_2$ but includes corrections for \rev{disc} thickness \citep{Romeo:2011aa}. They find $Q\approx 1-3$, with some increase near the center due to rising $\kappa_R$. In two thirds of their galaxy sample, $Q$ is dominated by the cold gas. Finally, \citet{Hitschfeld:2009aa} estimate $Q \approx 2-4$ for the M51 galaxy, with smaller $Q$ in spiral arms and in the outer \rev{disc}. The total $Q$, calculated as a three-component sum assuming equal velocity dispersions, shows no obvious radial trend but the gaseous component $Q_{\rm g}$ alone increases towards the galaxy center.

In summary, observations of low-redshift galaxies indicate marginally stable \rev{discs}, with occasional collapsing regions due to spiral arms or other gravitational perturbations. In contrast, high-redshift galaxies show more unstable regions and higher star formation rates. Our simulated galaxies resemble these high-redshift observations, however,  we do not find the values of $Q$ increasing towards the galaxy center. Our distribution of $Q$ is very patchy, and in general $Q$ increases towards the outer parts with low gas density.

Various numerical simulations of galaxy formation have also investigated the Toomre stability criterion. For isolated \rev{disc} galaxies, \citet{Li:2005aa} used the full \citet{Rafikov:2001aa} definition of the $Q$ parameter, and found that the star-formation timescale increases exponentially with $Q_R$. \citet{Li:2006aa} further found an anti-correlation between the SFR and the minimum value of $Q$ within the \rev{disc}. This trend probably arises because both quantities depend on the gas density: SFR$\;\propto \Sigma$ and $Q_{\rm min} \propto \Sigma^{-1}$. \citet{Westfall:2014aa} find a similar anti-correlation in their sample between the SFR surface density and $Q_{\rm min}$, although with large scatter. For our galaxies, the lowest value of $Q$ in patches is not representative of all star formation, but we checked that the $\Hm$ mass-weighted median value of $Q$ does not correlate with the SFR.

In simulations of high-redshift ($z\simeq2.3$) galaxies, \citet{Ceverino:2010aa} calculated a two-component $Q$ and found unstable regions in spiral arms and dense clumps. More recently, \citet{Inoue:2016aa} calculated the $Q_2$ parameter for their high-redshift clumpy \rev{disc} galaxies after removing the bulge and treating stars younger than 100~Myr as a gas component. They found relatively high values $Q\gtrsim 2-3$ in interclump regions and $Q<1$ only in very dense clumps. They also found that clumps begin forming with a high value of $Q$, which then decreases as clumps become denser. Such stable \rev{discs} could be a consequence of high mass concentration and low gas density, resulting from insufficiently strong stellar feedback. We find similarly high $Q$ values, low gas fraction, and high SFR$\gtrsim10\Msun\,{\rm yr}^{-1}$ in our weak feedback SFR50-SNR3 run.

Analysis of the FIRE simulations presented in \citet{Oklopcic:2017aa} has a very similar setup to ours: a 10~kpc square grid, with a 50~pc cell size smoothed to $\sim120$~pc to identify stellar clumps. They calculate the gaseous parameter $Q_g$ by approximating $\sigma={\sigma}_z$ and $\kappa=\Omega$, and find that many gas clumps at the $z\approx2$ output overlap with regions of $Q<1$, but do not match exactly. This is similar to our results that low values of $Q$ trace high-gas-density regions. We are also in agreement that the spatial coincidence between gas density peaks and SFR peaks washes out with increasing age of the stars.

To summarize our analysis of \rev{disc} stability, we can ask: How well does the $Q<1$ criterion work to predict the location and amount of star formation in these high-redshift galaxies? Is it a better criterion than a simple threshold on the $\Hm$ density? We think the answer is yes. 

\rev{Most of the neutral gas in our galaxies has values of $Q$ greater than one. However, even in this case turbulent discs may be unstable to gravitational collapse on small scales, below $\lambda_T$ \citep[e.g.,][]{Romeo:2010aa,Hoffmann:2012aa}. One way to evaluate the correspondence of the $Q$ criterion to star formation is to compare the amount of gas mass contained within a given threshold of $Q$ and SFR.}

The amount of $\Hm$ currently contained within the contours enclosing 99\% of SFR, averaged over 50 Myr, is similar to that in patches with $Q<1$. However, when we look only at regions containing 99\% of SFR within the shorter timescale of 10 Myr, the $\Hm$ mass shows much more variation among the different runs (with the same stronger-feedback prescription): from 0.04 to 90 times the mass of stars formed in this period. Even if we select the very narrow part of the \rev{disc}, calculating the column density of gas and SFR within only $\pm 0.2$~kpc, the range of variation still extends from 0.3 to 38 times. In contrast, the $\Hm$ mass selected by the $Q<1$ criterion varies only between 5.5 and 220 times the young star mass. This shorter range of variation makes the $Q$ criterion more useful for predicting future star formation.

Another way to make this comparison is to define a threshold in $\Hm$ surface density such that the mass contained in patches above that threshold matches the mass in patches with $Q<1$. We find that this threshold would vary from 48 to $110\Mpcs$ for the different strong-feedback runs, that is, by more than factor of two. For $\Hn$ gas the corresponding thresholds would be even higher: $140-250\Mpcs$. (For the weak-feedback run the thresholds are another order of magnitude larger.) These densities significantly exceed the prediction of \citet{schaye2004} model, in which transition to star-forming gas in present-day galaxies is expected to happen at $3-10\Mpcs$. This may be an evidence for denser and more compact ISM in galaxies at $z\approx 1-2$.

\begin{figure}
  \centering
  \includegraphics[width=\columnwidth]{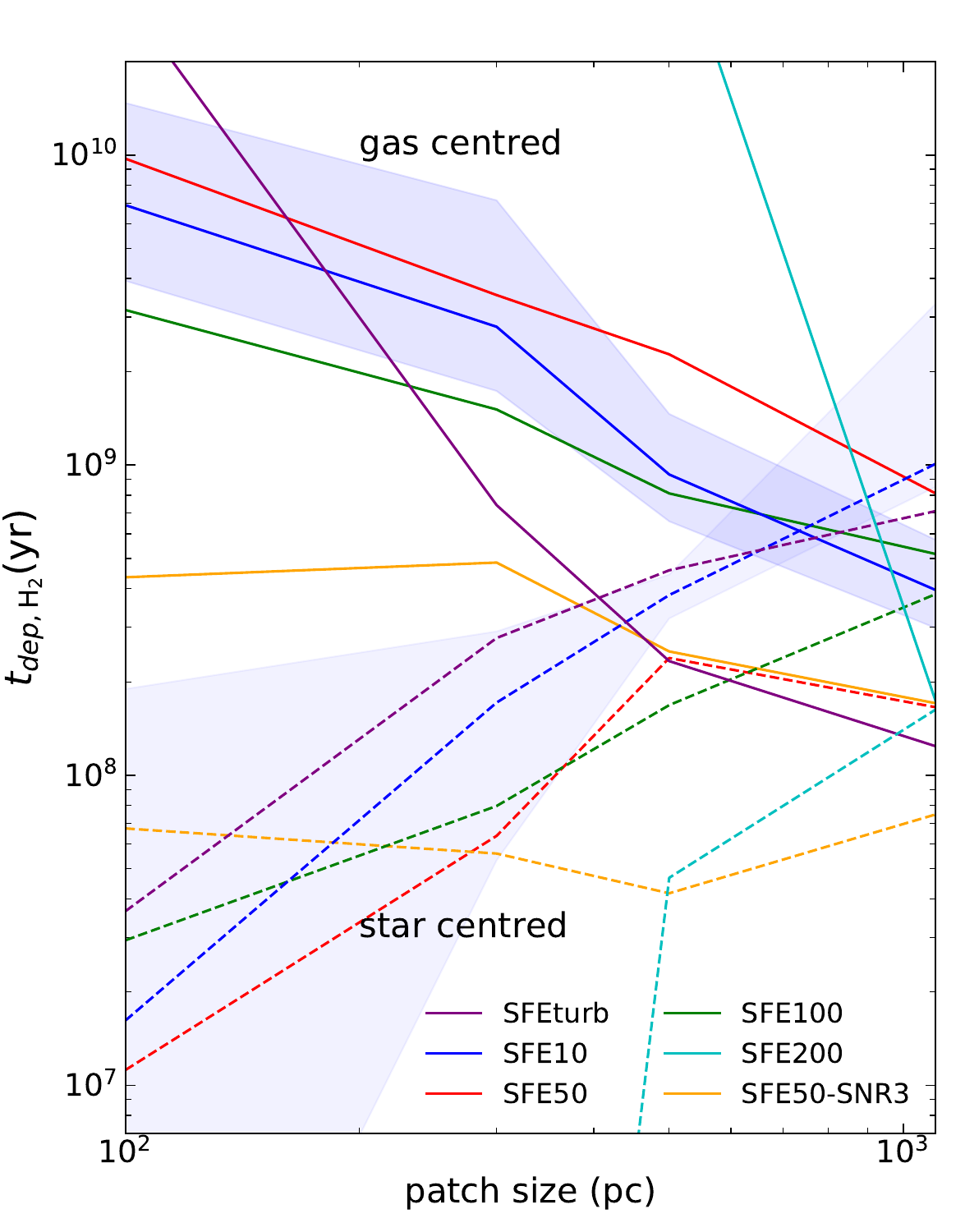}
  \vspace{-5mm}
\caption{Depletion time of $\Hm$ for "centered-on-gas" (solid lines) and "centered-on-stars" versions (dashed lines) as a function of patch size. For "centered-on-gas" version, we take the median of $\tdep$ in patches with $M_{\Hm}>10^6\Msun$. For "centered-on-stars" version, we take the median of $\tdep$ in patches with non-zero SFR. Shaded regions show the 40\%-60\% range of the values of $\tdep$ for SFE10 run, to illustrate the typical wide spread of the distribution.}
   \label{fig:tdep}
\end{figure}

\subsection{Depletion time}
\label{sec:discussion_depletion}

\citet{Utomo:2017aa} and \citet{Colombo:2018aa} measured the depletion time of molecular gas on kpc scale in the EDGE-CALIFA survey of nearby galaxies. They found $\tdep\approx 2.4$~Gyr, with large scatter of about 0.5~dex. The depletion time decreases near the center in some of the galaxies, especially at lowest masses $M_* \lesssim 10^{10}\Msun$. 

For high-redshift galaxies the depletion time of molecular gas appears to be shorter. According to \citet{Genzel:2010aa}, the depletion time for normal star-forming galaxies decreases from 1.5~Gyr at $z\approx 0$ to 0.5~Gyr at $z\approx 2$. \citet{Tacconi:2010aa, Tacconi:2013aa} measured $\tdep\approx 0.7$~Gyr, with a dispersion of 0.24~dex, in a survey of $z\approx 1-3$ galaxies with $M_*>2.5\times10^{10}\Msun$ and SFR$\gtrsim 30\Msun$~yr$^{-1}$. However, it is important to note the large scatter associated with this mean trend -- a small selection of galaxies may deviate significantly. For example, \citet{Tadaki:2018aa} find a very short depletion time of $10^8$~yr in a sub-mm starburst galaxy at $z=4.3$. This galaxy shows giant kpc-scale molecular gas clumps with low $Q\approx 0.3$, due to the very high gas density. \citet{Forrest:2018aa} also find spread of two orders of magnitude for the specific SFR of $1<z<4$ galaxies in the large ZFOURGE survey.

Our simulated galaxies, except in the weaker feedback SFE50-SNR3 run and the higher-redshift SFEturb output, show comparable values of $\tdep$ to the observations of $z\approx 1-2$ galaxies. When averaged in cylindrical shells, the depletion time roughly follows the trend of decreasing towards the center but given the strong patch-to-patch variation we do not study it further.

Recent observations by \citet{Rebolledo:2015aa} and \citet{Leroy:2017aa} have been able to probe individual star-forming regions on scales below 100~pc in nearby galaxies. They reveal a wide scatter of SFR density by three orders of magnitude at $\Hm$ densities $10-10^3\Mpcs$. Our simulations show a correspondingly large scatter in the local depletion times. Taking the patches above our adopted threshold on the $\Hm$ mass, which corresponds to the surface density of $10^6\Msun/(200\,\mathrm{pc})^2 = 25\Mpcs$, we find few individual patches with $\tdep < 10^8$~yr in the stronger-feedback runs. This corresponds to the lowest bound derived by \citet{Rebolledo:2015aa}. We find some patches with $\tdep > 10^{10}$~yr, which fall above the observed upper limit. However, such regions are more likely to escape detection because of their lower SFR. There is no systematic trend with $\epsff$ used in the simulation. Also, as with most of our results, the weaker-feedback run is an exception, as it contains patches with the depletion times as short as $10^7$~yr.

The high-redshift galaxies appear to have a high molecular gas fraction: 
\citet{Tacconi:2013aa} measured $f_{\rm mol}=M_{\rm mol}/(M_{\rm mol}+\Ms) \sim50\%$. After correcting for incompleteness, $f_{\rm mol}$ lowers to $30-40\%$. That fraction is larger than what we find for our simulated galaxies (about 15\%), even though our galaxies are less massive and therefore should be more gas-rich. \citet{Daddi:2010aa} also found the molecular gas fraction of galaxies at $z\approx 1.5$ to be $\sim50-65\%$. At very high redshift, the molecular gas fraction is even larger: \citet{Dessauges-Zavadsky:2017aa} measured $f_{\rm mol}\approx 60-79\%$ for a lensed $\Ms\sim 5\times10^9\Msun$ galaxy at $z\approx3.6$. In our galaxies the fraction of neutral gas reaches about 50\%, but the fraction of molecular gas stays low regardless of the value of local star formation efficiency adopted in the simulation (see Table~\ref{tab:basic}). 

Theoretical models predict the scaling of the depletion time with properties of star-forming regions. \citet{Semenov:2017aa} constructed an analytical model based on the mass conservation and the physical picture of rapid gas evolution between star-forming and non-star-forming states to study $\tdep$ and the fraction of gas that participates in star formation. They tested this model with a suite of $L_*$-sized galaxy simulations \citep{Semenov:2018aa} with different values of $\epsff$, feedback strength $f_{\rm boost}$ ($b$ in their notation), and star formation threshold. According to their model, gas regulation in galaxies is divided into two regimes: the self-regulation regime where feedback is strong or $\epsff$ is large enough, and the dynamics-regulation regime where feedback is weak or $\epsff$ is small. In the dynamics-regulation regime, the supply of star-forming gas is balanced by dispersal due to dynamical processes such as turbulent shear, differential rotation, etc. The depletion time is inversely proportional to $\epsff$, and the star-forming gas fraction is insensitive to $\epsff$ or feedback strength. In the self-regulation regime, gas spends most of the time in non-star-forming stages, and gas regulation is mainly controlled by star formation and feedback. The depletion time scales with feedback strength, but is insensitive to $\epsff$. The star-forming gas fraction is small and scales inversely with $f_{\rm boost}$ and $\epsff$. Although the model is formulated for the depletion time of all gas on kpc scales, the behavior of $t_{\rm dep,\Hm}$ and $\Hm$ fraction with different $\epsff$ and feedback strength is similar.

Our simulated galaxies fall in the self-regulation regime. Consistent with the \citet{Semenov:2017aa} model, our $t_{\rm dep,\Hm}$ on the largest scale shows a slightly decreasing trend with $\epsff$, although non-monotonic and with large scatter. It may be mainly due to the  fraction of gas in molecular phase, $M_\Hm/M_\Hn$, generally falling with $\epsff$. The weaker feedback run has slightly smaller $t_{\rm dep,\Hm}$ and higher $\Hm$ fraction than the other runs, which is also consistent with their model. The scale dependence of $\tdep$ is also similar, but our results are less regular and show significant scatter.

Galactic star formation relations are expected to break down below a certain spatial scale due to incomplete sampling of star formation or gas tracers, and relative motion of dense gas and young stars because of stellar feedback. The scatter of star formation relations, coming from the discreteness and stochasticity of star formation, and drifting of young stars, becomes more significant on smaller scales \citep{Feldmann:2011aa,Feldmann:2012aa}. \citet{Kruijssen:2014aa} constructed a model to describe this breakdown of SF relations based on the concept that a galaxy consists of many independent star-forming regions separated by some length scale, and that these regions are going through the SF process during which gas and/or young stars can be observed by some tracer. The timescale of the whole SF process is a combination of the epoch when gas is visible ($t_{\rm gas}$) and when stars are visible ($t_{\rm star}$), with some overlap time ($t_{\rm over})$. The scatter in $\tdep$ increases from $\sim 0.1$ dex on kpc scale to $\sim 1$ dex on tens of pc scale for randomly positioned apertures. Centering apertures on gas or stellar peaks systematically biases $\tdep$ -- centering on gas peaks overestimates $\tdep$ and centering on stellar peaks underestimates $\tdep$ -- making the "tuning fork" diagram, as shown in our Figure~\ref{fig:tdep}. The relative durations of the various phases of SF process ultimately determine the excess or deficit of $\tdep$ on small scales.

\citet{Kruijssen:2018aa} provide a detailed method to reconstruct the timescales of star formation and feedback ($t_{\rm gas}$, $t_{\rm star}$, $t_{\rm over}$) from the maps of gas and stellar flux. Using their method requires that the SFR for both gas-centered and star-centered apertures is averaged over the same time span. We find the ratio of the gas-centered to star-centered $\tdep$ about a factor of 100 at the smallest scale of 100~pc, with a very large variation between the different runs; these numbers exceed even the largest expected spread shown in \citet{Kruijssen:2018aa}. From our Figure~\ref{fig:tdep} we can at least see that our $t_{\rm gas}$ is much larger than $t_{\rm star}$ (taken to be 10~Myr here). From Figure~\ref{fig:H2map} we can infer that our $t_{\rm over}$ is definitely smaller than 50~Myr, and probably close to 10~Myr, which explains why we have so many infinities in the gas-centered determination of $\tdep$. The depletion time is so short on small scales when centered on stars that dense gas does not coincide with young stars, and causes formally infinite $\tdep$.

\section{Summary} \label{sec:summary}

We have investigated the structure of high-redshift ($z\approx 1.5-2$) galaxies in a suite of cosmological simulations with different star formation efficiency and feedback strength. Our main results are summarized below:
 
\begin{itemize}

\item Unlike the regular appearance of low-redshift \rev{disc} galaxies, the galaxies in our simulations have thick stellar \revis{components} with irregular, prolate shapes. The kinematics are dominated by turbulent motions and not by rotation. The stellar surface density profiles are approximately exponential, with the scale length of about 1~kpc. 

\item Although the vertical scale heights for all gas and all stars are large, cold molecular gas is concentrated \revis{to a relatively thin plane}. Young stars, which form from the molecular gas, likewise have the distribution with axis ratios $c/a = 0.1-0.2$.

\item Spatial correlation between the peaks of gas density and SFR deteriorates with the age of stellar population and almost disappears after $\sim 50$~Myr, because of stellar feedback dispersing old gas clouds around star-forming regions.

\item We calculate the maps of Toomre $Q$ parameter in patches of 200~pc, combining three components with different velocity dispersions: stars, molecular gas, and atomic gas. The median value of $Q$ weighted by $\Hm$ mass is in the range $0.5-1$, surprisingly close to unity given the irregular structure of the galaxies. 

\item The median value weighted by neural $\Hn$ mass is higher: $Q\approx1.5-2.6$. The $Q$ parameter in the weaker feedback run SFE50-SNR3 is systematically larger than in the other runs, because of the low gas density and high central mass concentration.

\item The dynamic range of $Q$ maps is much smaller than that of the $\Hm$ surface density maps, making the Toomre $Q$ parameter a better indicator of unstable regions that would collapse and form stars. The $Q$ parameter also depends on the spatial scale over which it is calculated: enlarging the averaging scale increases the value of $Q$. 

\item The depletion time of molecular gas in our galaxies is around 1~Gyr on the kpc scale, with large scatter from run to run. On smaller scales, $\tdep$ splits to systematically larger or smaller values when centering the aperture on gas peaks or stellar peaks, respectively. 

\end{itemize}

\section*{Acknowledgements}

We thank Miroslava Dessauges-Zavadsky, Andrey Kravtsov, Nir Mandelker, Roman Rafikov, and Vadim Semenov for useful discussions. 
This work was supported in part by NSF through grant 1412144.




\bibliographystyle{mnras}
\bibliography{gc,bzd} 




\bsp	
\label{lastpage}
\end{document}